\documentclass[fleqn,usenatbib]{mnras}

\usepackage[T1]{fontenc}

\DeclareRobustCommand{\VAN}[3]{#2}
\let\VANthebibliography\thebibliography
\def\thebibliography{\DeclareRobustCommand{\VAN}[3]{##3}\VANthebibliography}

\usepackage{graphicx}	
\usepackage{amsmath}	
\usepackage{amssymb}	
\usepackage{subcaption}
\usepackage{multirow}
\usepackage{array}
\usepackage{ctable}
\usepackage{multicol}
\usepackage{makecell}

\newcommand{\kepler}{{\it Kepler}~}

\newcommand{\re}{$R_{\oplus}$}

\usepackage{newtxtext,newtxmath}

\title[The SATCHEL pipeline]{The SATCHEL pipeline:  A general tool for data classified through citizen science}

\author[E. J. Safron et al.]{
E. J. Safron,$^{1}$\thanks{E-mail: ejsafron@gmail.com}
T. S. Boyajian,$^{1}$
and N. Eisner$^{2}$
\\
$^{1}$Department of Physics \& Astronomy, Louisiana State University, 202 Nicholson Hall, Baton Rouge, LA 70803, USA\\
$^{2}$Department of Physics, University of Oxford, Parks Rd, Oxford OX1 3PJ, United Kingdom\\
}

\date{Accepted XXX. Received YYY; in original form ZZZ}

\pubyear{2021}

\begin{document}
\label{firstpage}
\pagerange{\pageref{firstpage}--\pageref{lastpage}}
\maketitle

\begin{abstract}
Citizen science is a powerful analysis tool, capable of processing large amounts of data in a very short time.  To bridge the gap between classification data products from web-based citizen science platforms to statistically robust signal significance scores, we present the Search Algorithm for Transits in the Citizen science Hunt for Exoplanets in Lightcurves (SATCHEL) pipeline.  This open source, customizable pipeline was constructed to identify and assign significance estimates to one-dimensional features marked by volunteers.  We describe the functional capabilities of the SATCHEL pipeline through application to features in photometric time-series data from the {\it Kepler} Space Telescope, classified by volunteers as part of the Planet Hunters citizen science project hosted on the Zooniverse platform.  We evaluate the SATCHEL pipeline's overall performance based on recovery of known signals (both simulations and signals corresponding to official {\it Kepler} Objects of Interest) and relative contamination by spurious features.  We find that, for a range of pipeline hyperparameters and with a reasonable score cutoff, SATCHEL is able to recover volunteer identifications of over 98\% of signals from simulations corresponding to exoplanets $> 2~R_\oplus$ in radius and about 85\% of signals corresponding to the same size range of KOIs.  SATCHEL is transparently adaptable to other citizen science classification datasets, and available on GitHub.

\end{abstract}

\begin{keywords}
methods: data analysis -- software: documentation -- software: public release -- planets and satellites: detection
\end{keywords}



\section{Introduction}

Large-scale scientific experiments and surveys are producing ever more vast amounts of data.  The \kepler Space Telescope, launched in 2009 \citep{bib.boru2010}, produced about thirty gigabytes of data per month \citep{bib.chri2021}.  By contrast, its currently-operating successor, the Transiting Exoplanet Survey Satellite \citep[{\it TESS},][]{bib.rick2015}, generates more than twice that in a single day.  The upcoming Rubin Observatory \citep[formerly the Large Synoptic Survey Telescope, LSST,][]{bib.juri2017} will generate a colossal twenty terabytes of raw data per night.  For a single scientist or small team of scientists, it is impractical at best and impossible at worst to analyze such copious data.  For some datasets, automated or machine learning algorithms have been shown to be successful at various types of tasks \citep{bib.baro2019}, and the field of machine learning is evolving rapidly in response to the rising need for innovative methodologies.

For analysis tasks involving pattern recognition, however, many automated techniques still fall short of the remarkable capabilities of the human brain.  This has led to the development and increasingly widespread use of ``citizen science,'' in which the general public engages in scientific research by contributing intellectual effort, knowledge, or tools and resources \citep{bib.gord2013}.  One popular platform that offers volunteers the opportunity to assist in data analysis\footnote{While not the focus of this work, volunteer participation can also come in the form of data collection, as exemplified by the plethora of \href{https://www.birds.cornell.edu/citizenscience}{citizen science projects hosted by the Cornell Lab of Ornithology} \citep{bib.bonn2009}, and by the wealth of astronomical measurements provided by amateurs as part of, for instance, the American Association of Variable Star Observers (AAVSO) or the TESS Follow-up Working Group.} is the Zooniverse\footnote{\href{http://zooniverse.org}{http://zooniverse.org}} \citep{bib.lint2008}, where volunteers (henceforth ``users'') can access curated scientific data (``subjects'') and help analyze them by performing simple tasks (``classifications'').  Through the Zooniverse, citizen science has been successful over the last decade for studies involving tasks such as:
\begin{itemize}
    \item Identification of novel objects, such as Hanny's Voorwerp \citep{bib.lint2009} and Boyajian's Star \citep{bib.boya2016},
    \item Categorization and cataloging \citep{bib.will2013,bib.john2015,bib.simm2017,bib.dick2018,bib.chri2018,bib.zink2019}, and
    \item Creation of training data sets for machine learning algorithms \citep{bib.feng2017}.\footnote{In addition to astronomy and physics, the Zooniverse also hosts projects that apply citizen science data analysis in the areas of climatology, biology, history, and various humanities.}
\end{itemize}
In addition to its versatility, citizen science accomplishes many of its project goals extremely quickly.  For example, users of the Snapshot Serengeti project fully processed 18 months of backlogged data within the first three days of the project's establishment on the Zooniverse platform \citep{bib.swan2015}.  In 2015, it was estimated that, on average across all Zooniverse projects, volunteers processed an amount of data equivalent to 34 years of full-time work by a single researcher throughout the course of a project \citep{bib.cox2015}.

Generally, a single subject is classified by many users, as the skill levels of different users may vary significantly.  Crowdsourcing, as this method is sometimes called, effectively suppresses bias from any one individual user, and it has been shown that for some kinds of classification tasks, the combined judgment of many volunteers is as good or better than that of a single expert \citep{bib.will2013}.  Methods for aggregating these classifications and determining the final verdict for a given subject vary according to the needs of the project and the nature of the data.  Some projects find it sufficient to rely on simple crowd majority voting \citep{bib.kuch2016,bib.zink2019}.  Others find it more beneficial to ``weight'' the votes according to some measure of the skill of each user who engaged with the subject.  User skill is typically measured based on
\begin{itemize}
    \item User performance on ``gold-standard'' data or simulations with known ``correct'' classifications \citep[e.g.,][]{bib.mars2016,bib.eisn2020b},
    \item User classification agreement with consensus \citep[e.g.,][]{bib.will2013,bib.john2015},
    \item A proxy for engagement level or attentiveness to detail \citep[e.g.,][]{bib.simp2012},
\end{itemize}
or a combination of two or more of the above metrics \citep[e.g.,][]{bib.schw2012,bib.simm2017}.  If consensus agreement is used to assign user weights, weighted vote fractions (henceforth, ``scores'') may be recalculated and user weights readjusted, repeating this process several times to achieve convergence of results.

In this paper, we present the Search Algorithm for Transits in the Citizen science Hunt for Exoplanets in Lightcurves (SATCHEL) pipeline, an open source,\footnote{The SATCHEL pipeline is publicly available on GitHub, at \href{https://github.com/ejsafron/PH-pipeline}{https://github.com/ejsafron/PH-pipeline}.} Python-based pipeline to process classifications comprised of arbitrary, one-dimensional markings made by citizen science volunteers.  By one-dimensional, we refer to user markings with boundaries along only one measurement axis, rather markings whose midpoint requires both $(x,y)$ pixel coordinates.  SATCHEL is designed to score individual features in classification subjects, rather than subjects themselves.  This is especially useful for projects in which multiple types of significant signals may be present in a single classification subject, prompting a need for the ability to filter out some types of features while retaining others.  SATCHEL employs a user weighting scheme based on both gold-standard data and consensus agreement, and the latter weighting process is iteratively coupled to the calculation of scores.  Unlike the weight-score coupling implemented in other analysis pipelines, SATCHEL does not execute a predetermined number of iterations to achieve convergence.  Instead, convergence is determined by stability of the user weight distribution.

In the following sections, SATCHEL is described via application to classification data from the Planet Hunters\footnote{\href{https://planethunters.org}{https://planethunters.org}} project \citep[PH,][]{bib.fisc2012}.  An overview of the PH project and interface is given in Section~\ref{sec.PH}.  The SATCHEL feature isolation, user weighting, and scoring processes are described in detail in Section~\ref{sec.methods}.  A performance study, including measurement of purity and completeness, is presented in Section~\ref{sec.recovery}.  Section~\ref{sec.discussion} contains a summary and discussion.  Appendices~\ref{sec.optimize},~\ref{sec.decay}, and~\ref{sec.scorecutoff} provide details of pipeline optimization.  Details regarding scientific motivation, sample selection criteria, and astrophysical implications of the study are deferred to an accompanying paper, hereafter ``Paper II.''

\section{Planet Hunters}\label{sec.PH}

The PH project, as its name suggests, is a citizen science project designed to allow volunteers to ``hunt'' for exoplanet signals by marking dip-like features in time-series stellar brightness data, which may correspond to a planetary companion passing in front of a star from our line of sight.  This ``transit method'' of exoplanet search has been extremely successful over the last decade, particularly since the launch of the {\it Kepler} Space Telescope \citep{bib.boru2010,bib.koch2010}, which increased the number of known exoplanets by more than a factor of ten over the course of its four-year main mission.\footnote{At the time of writing, the number of confirmed and candidate exoplanet detections attributed to \kepler is 4780, according to the NASA Exoplanet Archive.}  Despite its prolific success, however, the transit method has weaknesses and biases \citep{bib.yee2008,bib.kipp2016}.  For instance, even if a host-planet system has a size ratio and geometrical alignment favorable to transit detection with a given instrument, the planet may still not be detected if its orbital phase does not permit a transit during the time baseline of observations.  This becomes a significant obstacle for exoplanets at large separations from their hosts, as orbital period increases with separation.  Furthermore, many automated algorithms designed to search for transit signals in observational data require not one, but three transits to occur over the baseline of observations in order to be flagged as a potential exoplanet candidate.  While this practice does efficiently weed out some types of aperiodic signals that can otherwise mimic true transit signals, it also makes detecting exoplanets with orbital periods above a certain threshold explicitly impossible.

While a few innovative automated pipelines have been developed to circumvent the three-transit criterion \citep[e.g,][]{bib.fore2016,bib.herm2019}, visual inspection has also proven to be effective at picking out transit signals in time-series data \citep{bib.schw2012,bib.wang2013,bib.schm2014,bib.wang2015,bib.ueha2016,bib.eisn2020a,bib.eisn2020b}.  The PH project maximizes the advantage of such human capabilities by crowd-sourcing the task of finding transit signals and other interesting transient features \citep{bib.fisc2012}.  In 2018, PH was refurbished to become PH {\it TESS} (PHT), hosting only data from the {\it TESS} mission for classification.  Prior to that, however, the PH site hosted and classified \kepler data.  In particular, between 2013 and 2015, PH users classified all time-series data from the full, four-year time baseline for the coolest subset of the original \kepler targets, dwarf stars of the {\it M} spectral type.  These classifications are the focus of the application described throughout the remainder of this paper.

\subsection{Data and Interface}\label{sec.interface}

\begin{figure*}
\centering
\includegraphics[width=1.0\linewidth,height=6.8cm]{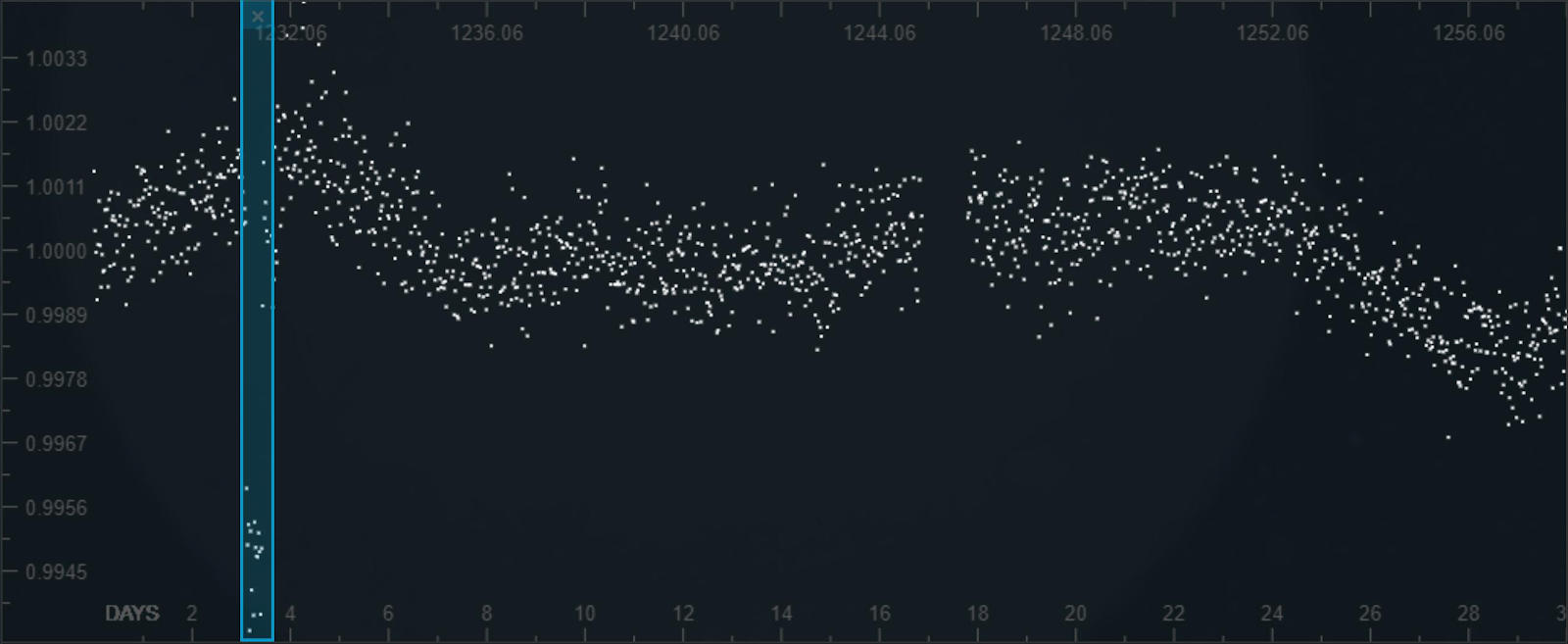}
\caption{Screenshot of the light curve subject display in the PH web interface, as it appeared in August 2017, showing a typical 30-day segment of \kepler data.  The flux is normalized on the vertical axis, and the horizontal axis is time, shifted to begin at day ``zero.'' Data are shown (without errors) by white points.  The user viewing the subject is able to highlight any transit-like dip features by placing, dragging, and adjusting blue boxes such as that shown here between days 3 and 4.\label{fig.PH}}
\end{figure*}

Data from the \kepler Space Telescope were transmitted and released in a series of 17 quarters, each consisting of about three months of high-precision photometric time-series, or {\it light curves}, for about 150,000 stars.  The data products released by the \kepler science team included both ``long-cadence'' light curves, composed of photometric observations taken every $29.4$ minutes, and ``short-cadence'' light curves, in which observations were taken every $58.89$ seconds.  The raw long-cadence light curves were scaled and corrected by Pre-Search Data Conditioning (PDC) in the \kepler science processing pipeline \citep{bib.jenk2010} to compensate for systematic instrumental errors, and both long- and short-timescale spurious and non-astrophysical trends.

For display in the classification interface on the PH website, the PDC-corrected light curves from each quarter were obtained for 5625 likely M-dwarf targets\footnote{Paper II discusses in more detail how this sample selection took place, as well as the post-classification pruning of the sample from 5625 stars down to the 3262 whose PH classification data function as the input for this work.} by the PH team from the \kepler Mission Archive\footnote{\href{http://archive.stsci.edu/kepler}{http://archive.stsci.edu/kepler}} hosted by the Mikulski Archive for Space Telescopes\footnote{\href{http://archive.stsci.edu}{http://archive.stsci.edu}} (MAST).  Each quarter-long light curve was divided into three segments (usually about 30 days each), and then normalized by the mean flux of the segment, as shown in Fig.~\ref{fig.PH}.  These light curve segments were displayed in the PH user interface as classification subjects, ordered randomly and devoid of any indicator of which star the data represented.  The only metadata available to a user were the subject's effective temperature, magnitude, and estimated radius.

\begin{figure*}

\begin{subfigure}[b]{.48\linewidth}
\includegraphics[width=\linewidth]{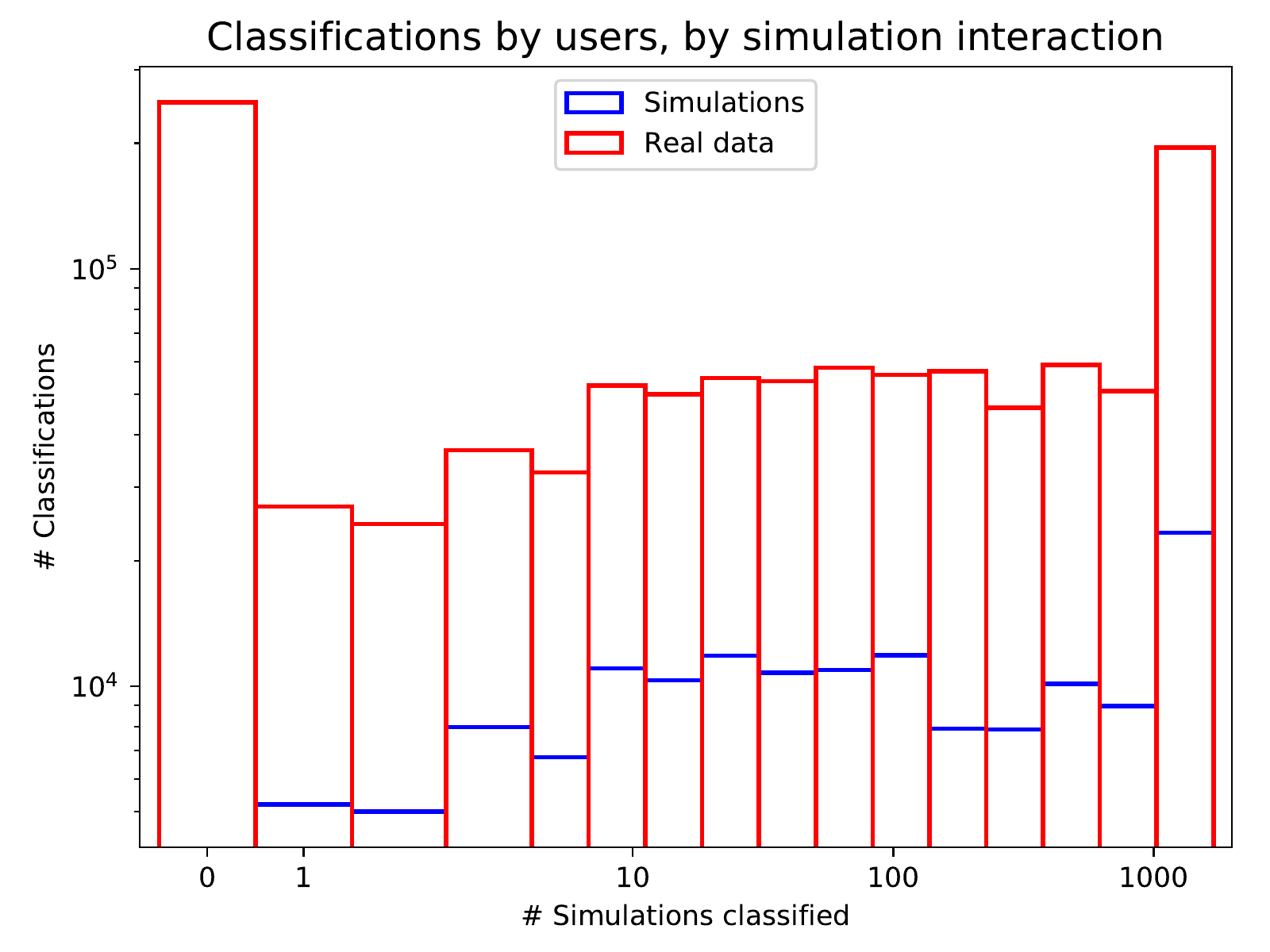}
\caption{}\label{fig.classsynthvsreal}
\end{subfigure}
\begin{subfigure}[b]{.48\linewidth}
\includegraphics[width=\linewidth]{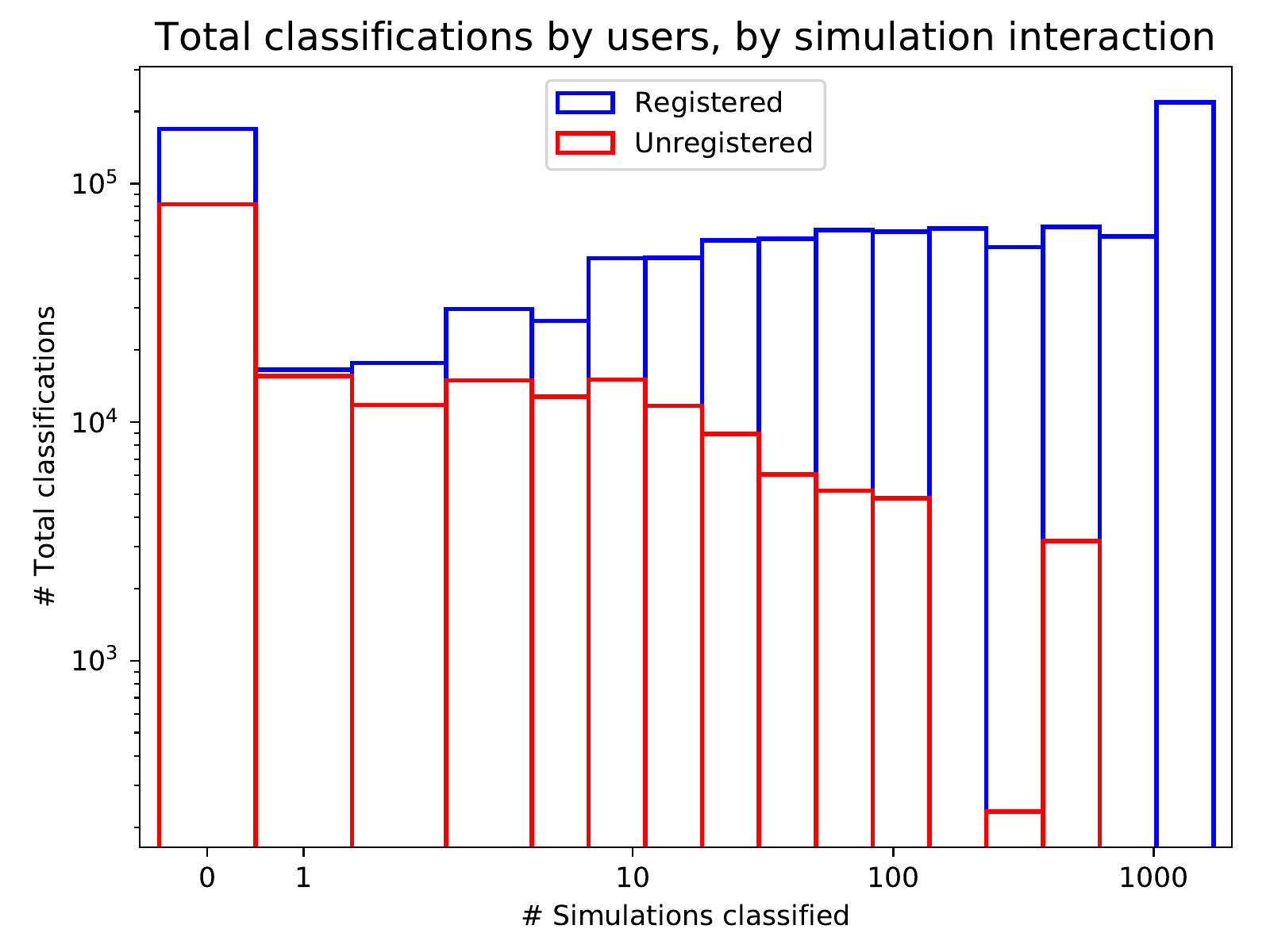}
\caption{}\label{fig.classsynthinter}
\end{subfigure}

\begin{subfigure}[b]{.48\linewidth}
\includegraphics[width=\linewidth]{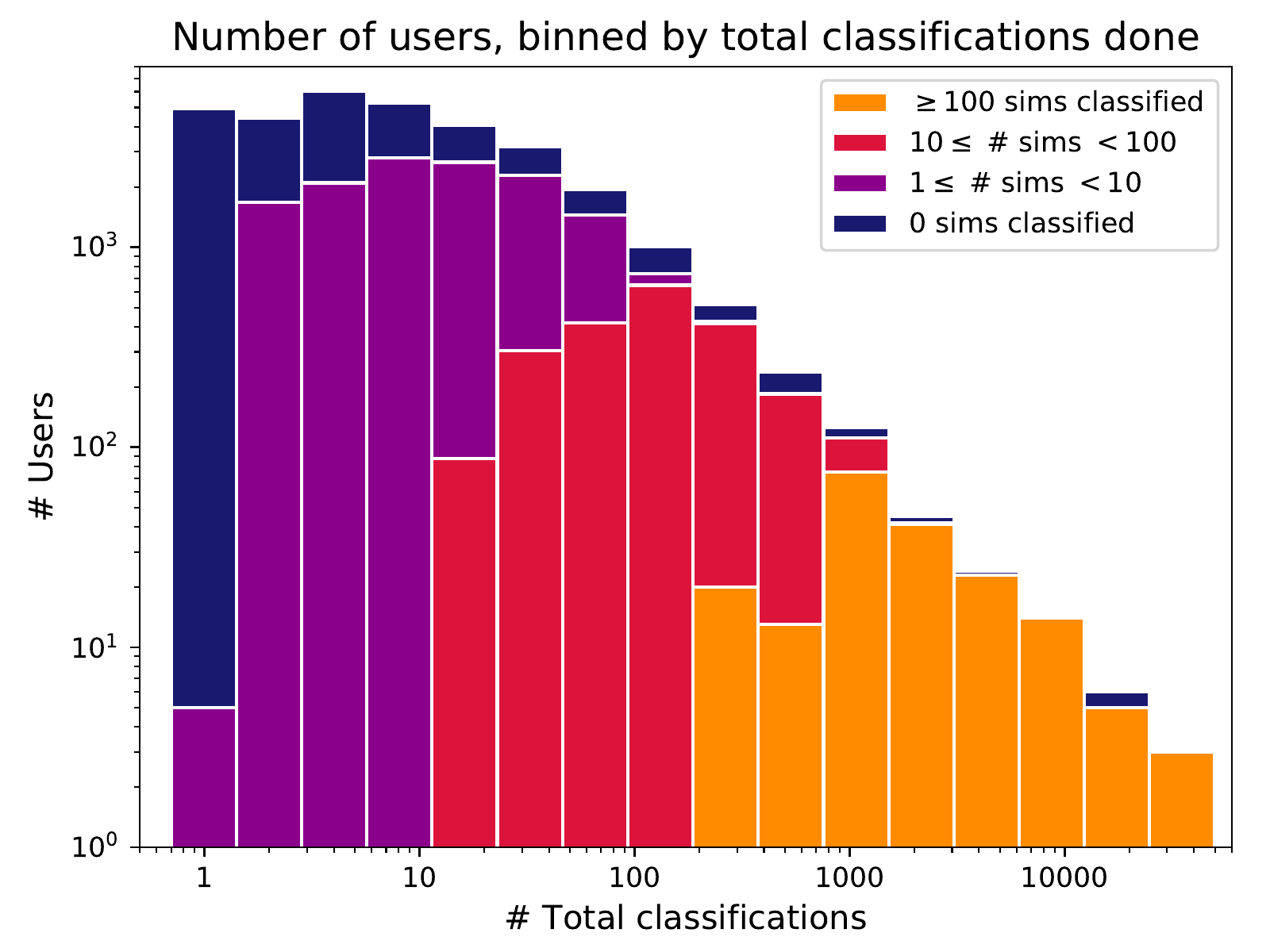}
\caption{}\label{fig.usersbyclass}
\end{subfigure}
\begin{subfigure}[b]{.48\linewidth}
\includegraphics[width=\linewidth]{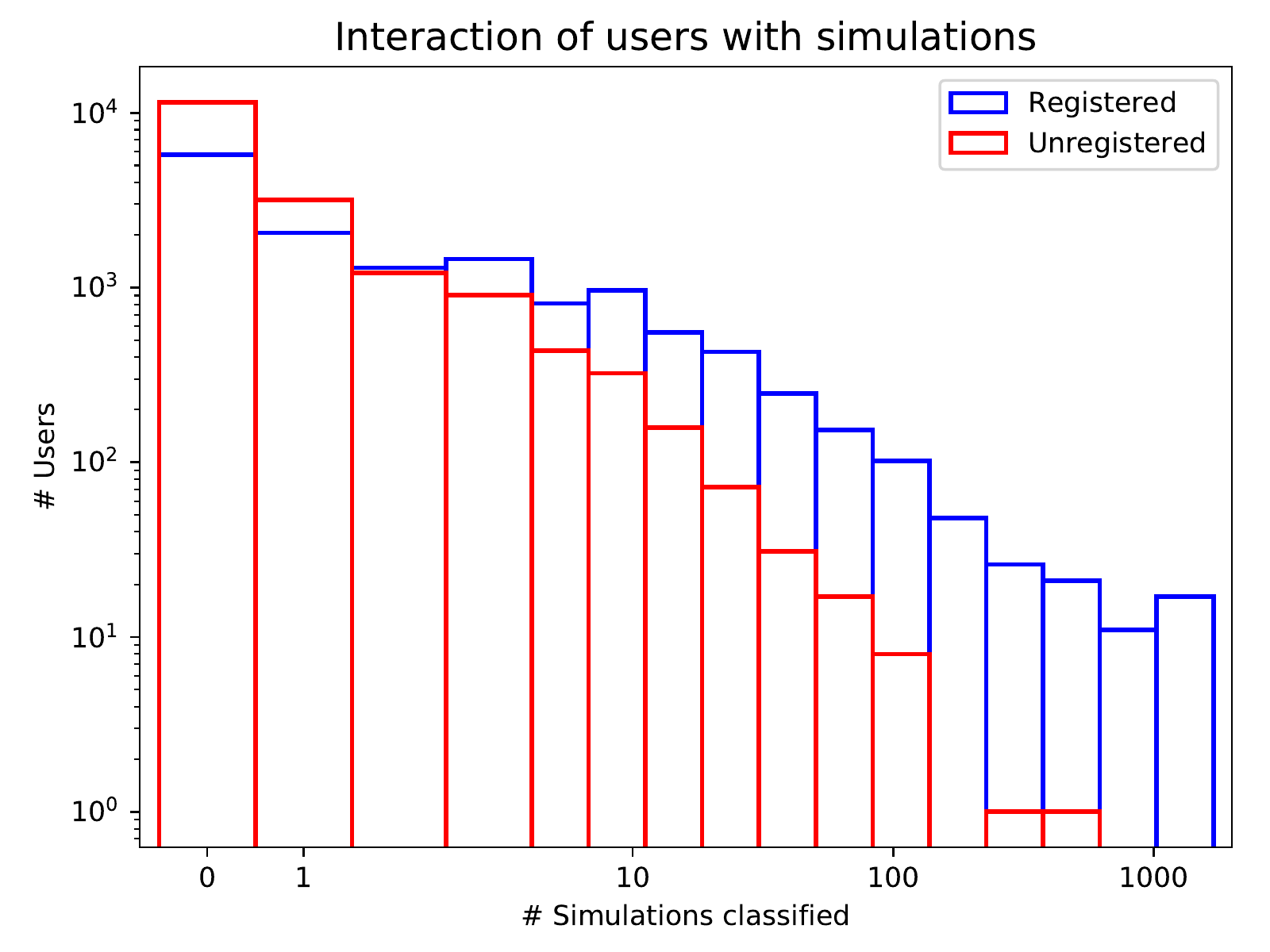}
\caption{}\label{fig.synthinter}
\end{subfigure}

\caption{Several different methods of gauging user interaction with the PH site, including registered vs. unregistered users and classifications of simulations vs. real data.  Note, from Fig.~\ref{fig.classsynthvsreal}, that over 200,000 classifications of real data were done by users who had never seen simulations, and more of those classifications were done by registered users than unregistered, as shown in Fig.~\ref{fig.classsynthinter}.  Fig.~\ref{fig.usersbyclass} emphasises that some individual users classified hundreds or even thousands of subjects without seeing a single simulation.  And Fig.~\ref{fig.synthinter} shows explicitly the large number of registered users who did not see any simulations.}
\label{fig.userinteraction}
\end{figure*}

The first time a user accessed the PH site, a brief tutorial was shown in a pop-out interface to guide the user through classification of a typical light curve.  An exoplanetary transit example was shown as a localized, dip-like feature, on the order of a couple percent in brightness measurement.  The user was instructed to draw a box marking the beginning and end of the transit feature, and do the same for any other transit features on the light curve.  After a brief confirmation dialogue, the user was shown real data to classify.  Users were given the option to register with a Zooniverse account, or login to their existing account if they already had one.  An unregistered user was reminded of this option periodically if they remained on the site to classify more than one light curve.  If an already-registered user logged in to the site, they would not be shown the tutorial again, but the tutorial was always accessible via the menu bar.  About $43.9$\% of users were registered, performing $84.2$\% of all classifications (see Fig.~\ref{fig.classsynthinter}).

Once a light curve subject was independently classified by ten different users, it was retired from the random display rotation.  This number was chosen based on calculations from early results of the Galaxy Zoo\footnote{\href{https://www.zooniverse.org/projects/zookeeper/galaxy-zoo/}{https://galaxyzoo.org}} citizen science project \citep{bib.lint2008}, which showed that an 80\% user agreement out of 10 classifications was equivalent to a 5$\sigma$ detection.

Users were also shown simulations, constructed by the PH science team using real \kepler light curves injected with synthetic transit signals.  The shape of each synthetic signal was generated based on parameters pulled at random from a range of exoplanet radius and orbital period values (further details in Section~\ref{sec.simreco}.  All simulations contained at least one injected signal.  Initially, 40 \kepler targets with effective temperatures $T_\text{eff}<4200$ and a representative variety of stellar radii, $K$-band magnitude, and activity level were chosen as simulation hosts.  Combinations of hosts and synthetic transit parameters were used to create 4000 30-day simulation light curves.  Of the 40 initially-chosen target stars, 24 remained in our M-dwarf sample after filtering out likely giant stars \citep{bib.gaid2016,bib.berg2020} (a process described in detail in Paper II), which left 2268 simulations.  And of these simulations, finally, more than half were determined to be too difficult for users to identify through visual inspection, due to the extreme shallowness of signals produced by very small planets.  Thus, only the 1428 simulation subjects containing synthetic transits from exoplanets larger than $1~R_\oplus$ were retained for classification and analysis.

The manner in which the simulations were presented was indistinguishable from that of the real, unaltered data, including the display of stellar property metadata.  Only after a user had classified the subject were they informed that it was a simulation, and shown where the synthetic transit signal(s) had been injected.  The 1428 simulation subjects were classified a total of $149,921$ times, with an average of 105 classifications each.  About $85.8$\% of these classifications were done by registered users.  The users who classified each simulation were split randomly into groups of ten.  For each of these sets of ten users, unique IDs were generated for the subject and its corresponding synthetic exoplanet and signals.  This resulted in an effective total of 15,233 simulations.  These simulations allow for both an individual user's performance and the recovery efficiency of PH as a system to be measured \citep[][hereafter S12]{bib.schw2012}, as discussed in Sec.~\ref{sec.simreco} and Appendices~\ref{sec.mfparam} and~\ref{sec.scorecutoff}.  

Figure~\ref{fig.userinteraction} provides several graphical metrics for gauging user interaction with the PH site, each of which offers its own insight into the user base and classifications used in this study.  Fig.~\ref{fig.classsynthvsreal}, by clearly displaying the colossal number of classifications done by users who saw no simulations, succinctly illustrates the necessity for developing a sophisticated weighting scheme capable of quantifying the performance of all users, including those who interacted with no simulations.  This is a significant departure from the methodology of the PH TESS analysis pipeline \citep{bib.eisn2020b}.  Fig.~\ref{fig.synthinter} shows, in a more general sense, how the users have interacted with the simulations, divided by color into registered and unregistered users.  The excess of registered users with high numbers of simulation classifications is not unexpected; rather, it's interesting that there are some users, while few, who enter the site and in one session classify enough light curves to have seen hundreds of simulations.  Fig.~\ref{fig.classsynthinter} complements this by showing the total number of classifications done by users in these same simulation classification count bins, again divided by color into registered and unregistered users.  Over all, though unregistered users outnumber registered users, the registered users classified significantly more data than the unregistered users, about $84.2$\%.  Even among these registered users, however, Fig.~\ref{fig.synthinter} shows that many did not classify any simulations.

Fig.~\ref{fig.usersbyclass} breaks down the user population in bins of total classifications done, stacked for visual clarity.  This view in particular gives some potential insight as to why the first bin of Fig.~\ref{fig.classsynthvsreal} is so large:  If a user visited the PH site and classified only one light curve, it was most likely not a simulation, as evidenced by the sim-to-real ratio of 5:4938 in the first bin of Fig.~\ref{fig.usersbyclass}.  In this plot, we also see the clear pattern of users who classify more subjects naturally being the ones who classify more simulations, but there are a few notable exceptions visible---namely, one user who somehow classified over ten thousand subjects without encountering a single simulation.  Investigation into this case is ongoing.

These simulations are an integral part of the user weighting process (see Sec.~\ref{sec.weighting}), as well as in determining the recovery efficiency of the PH system (Sec.~\ref{sec.recovery}), which is fundamentally necessary for calculating exoplanet occurrence rates in Paper II.

\subsection{User classification outputs}\label{sec.mdwarf-classifs}

There were 1,252,539 unique classifications of our sample, comprised of 1,856,573 markings and non-markings, made by 31,781 users between 18 September 2014 and 9 January 2016.

The ``raw data'' output by the PH backend was in the form of a \texttt{.csv} file containing all classifications done on all subjects by all users up to the date of the file export, including simulations.\footnote{The backend data exports for PH {\it TESS} are the same format, with only very minor differences.}  This file contained a row for every single user marking made (multiple markings made by the same user during a single classification of a subject constituted multiple rows), as well as rows indicating when a user saw a subject but chose not to mark anything.  A full itemization of fields delivered by this file is described on the SATCHEL GitHub, including which fields are necessary as input for the algorithm.  Of particular importance for the following discussion are:  the classification ID (unique to each instance of a user viewing a subject); a subject ID (unique to each 30-day light curve segment); the location string of the raw \texttt{.json} file on the PH backend where the subject data was stored; and the ``global x'' values that denoted the bounds of the user mark relative to the beginning of the \kepler mission time baseline, beginning at Barycentric Julian Date (BJD)$-2454833.0$ (or \texttt{NaN}s if no marking was made).

The following were also either extracted or derived from the location string for each row:   The \kepler Input Catalog (KIC) identifier of the target star to which the light curve subject corresponds; the \kepler ``type,'' indicating whether the target was observed during the main \kepler\ mission (``{\it K1}'') or the supplementary {\it K2} mission \citep{bib.osbo2016,bib.laco2018}, the latter of which were summarily filtered from our sample to leave only data from the \kepler prime mission; a Boolean indicating whether or not the subject is a simulation; and the synthetic exoplanet ID for simulations (or \texttt{NaN} if the subject is not a simulation).  Note that a single synthetic exoplanet may generate mutliple synthetic transit signals in a simulation subject, provided that it has an orbital period of less than about 15 days.  Thus, there are unique identifiers for both the synthetic exoplanet (included in the classification file) and identifiers for the individual synthetic transit signals (not included in the classification file, but recorded elsewhere; see Sec.~\ref{sec.upweighting}).

\section{Methods}\label{sec.methods}

For past analysis of the PH data, it was sufficient to calculate for each light curve subject a ``score,'' which measured the likelihood that the subject contained a significant feature (S12).  If we were to apply this technique to the \kepler M-dwarf PH classifications, we would expect many of the features responsible for high scoring subjects would be either short-period signals produced by KOIs found in previous studies or signals from eclipsing binary stars, the most common source of false detections in \kepler exoplanet searches.  Short-period KOIs provide a useful tool for measuring the recovery efficiency of the SATCHEL pipeline (see Sec.~\ref{sec.KOIreco}), as well as for comparing the results of our statistical occurrence rate study in Paper II to other contemporary results.  However, to find longer-period signals that are more difficult or impossible to find using standard automated searches, a straightforward way to filter known signals out of a subject while preserving a measure of the significance of remaining features is required.  Thus, rather than scoring each subject, the pipeline scores individual features.  S12 does not take this approach in their study; the PH {\it TESS} pipeline does, notably, score individual features, but using a very different weighting scheme in their algorithm \citep{bib.eisn2020b}.  In particular, \citet{bib.eisn2020b} do not measure the performance of users who have not classified simulations.

\begin{figure*}
\centering
\includegraphics[width=1.0\linewidth]{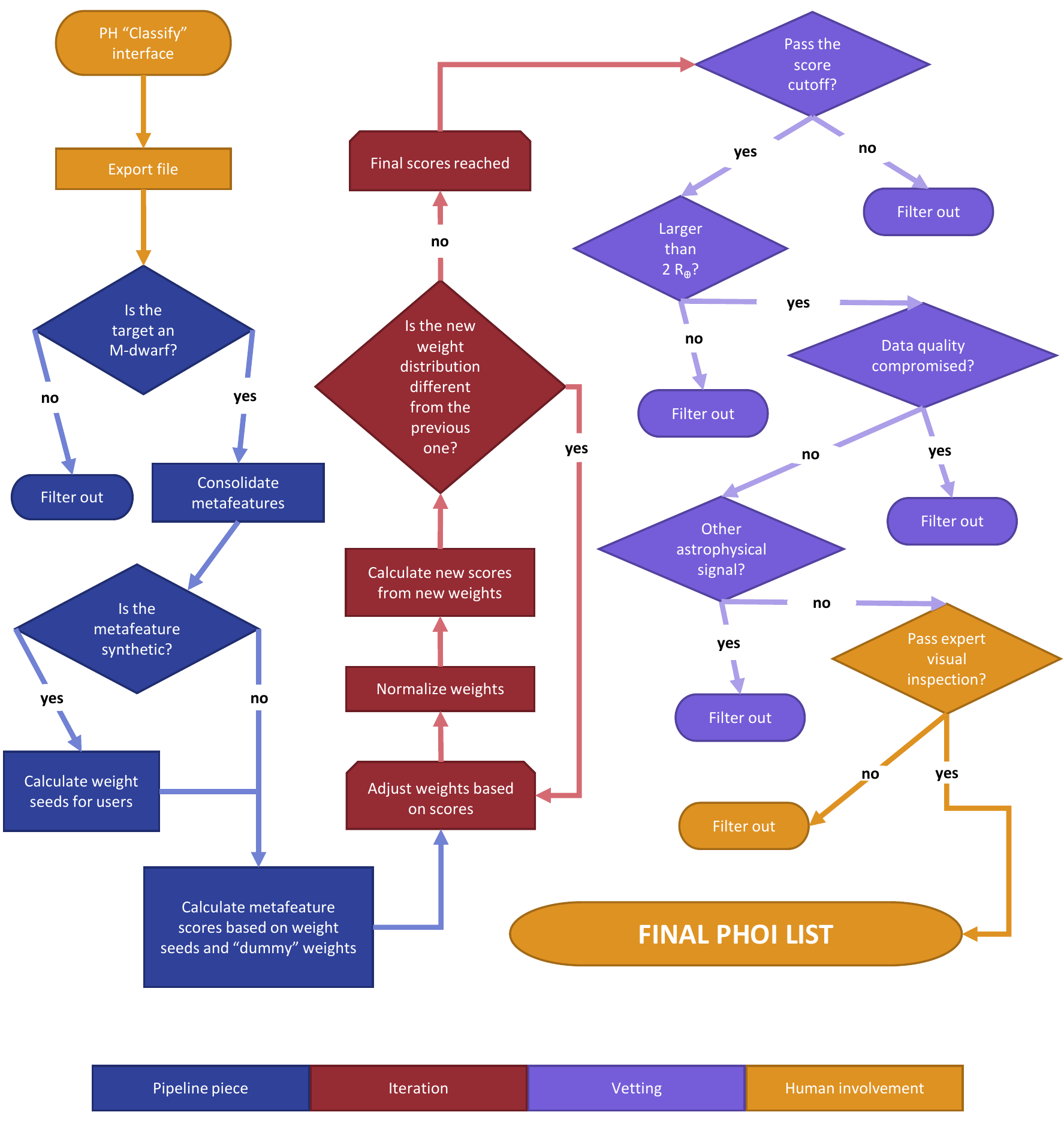}
\caption{Flowchart to illustrate the general process of the PH pipeline, along with the subsequent steps necessary to reach a final PHOI list.  Steps in orange bubbles represent steps of the process that require human involvement.  Blue and red steps are fully automated, the latter of which comprise the iterative portion of the SATCHEL pipeline.  Purple steps are vetting, which is described in Paper II.\label{fig.flowchart}}
\end{figure*}

The general structure of the SATCHEL pipeline is as follows.  Features of interest, which we call {\it metafeatures}, are identified by consolidating user marks that coincide on a given subject, as described in Sec.~\ref{sec.metafeatures}.  Then, following the strategy of S12, ``weights'' for users who classified simulations are assigned, based on skill at identifying synthetic transits (Sec.~\ref{sec.weighting}).  Using the resulting weights (and dummy weights equivalent to zero upweight and zero downweight for users who did not classify simulations), preliminary scores are calculated for all metafeatures (Sec.~\ref{sec.scoring}).  Then, using these initial distributions as ``seeds,'' the user weights and metafeature scores are iteratively coupled, as described in Sec.~\ref{sec.iteration}:  The weights for all users are adjusted based on user agreement with the majority on highly-scoring features, and scores are subsequently adjusted based on the new weight distribution.  The coupling is iterated until the user weight distribution converges with a stable mean weight of 1.0, from which the final metafeature scores are calculated.  Figure~\ref{fig.flowchart} illustrates how the individual pieces of the PH analysis pipeline are designed to work together, from parsing the data exported by the PH website backend up to obtaining a fully-vetted list of Planet Hunters Objects of Interest (PHOIs).  The following subsections detail each step of this process.

\subsection{Metafeatures}\label{sec.metafeatures}

The list of metafeatures to be scored is constructed by extracting features directly from the marks that users have made on the light curve subjects.  More obvious dip features are marked by multiple users, and these marks are consolidated to represent a single metafeature.  Thus, we do not score any feature more than once.  This consolidation is done using the pipeline piece named \texttt{build-metafeatures.py} (available in the ``aux'' folder of the SATCHEL GitHub).

Because SATCHEL is intended to be useful for assessing the significance of signals in data from other projects, which may have unique criteria designating which marks should or should not be ingored or how marks should be consolidated, the process of identifying metafeatures is considered technically auxiliary to the main pipeline.  The \kepler data on PH, for instance, contain many data gaps of varying duration, usually on the order of days, caused by such things as scheduled quarterly rolls of the spacecraft, sudden shifts to safe mode due to cosmic ray impacts, and spacecraft rotation toward Earth for data downlinking \citep{bib.stum2012}.  Users frequently mark these gaps, or mark anomalous features at the beginning or end of the gaps.  To filter out these marks, gaps in the \kepler light curves were identified by \texttt{NaN} entries for the PDCSAP flux.  Any user mark for which more than half the mark duration lies inside a gap was discarded.

\begin{figure}
\centering
\includegraphics[width=1.05\columnwidth]{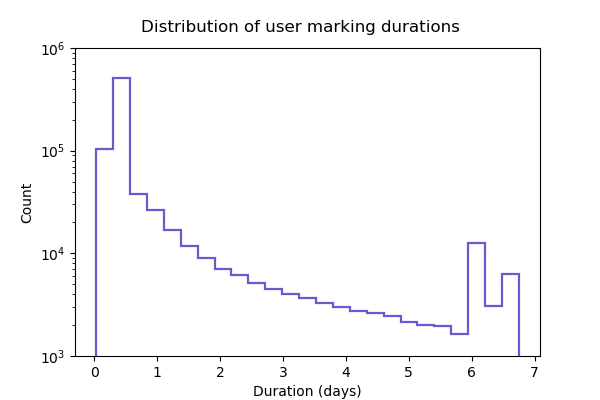}
\caption{Distribution of durations of user markings on non-synthetic M-dwarf lightcurves, from raw classification data (not metafeatures).  The maximum width users were allowed to draw in the PH interface was about 6.74~days.\label{fig.durations}}
\end{figure}

Very ``wide'' user marks were also discarded.  In the PH data, there are a significant number of these wide user marks, indicating dip features with very long durations (see Fig.~\ref{fig.durations}).  The duration of an exoplanet transit increases with both stellar radius and $\sqrt{a}$, where $a$ is the semi-major axis of the planet's orbit.  However, the stellar sample for this study has been filtered of giant stars (see Paper II), and the geometrical probability of observing a transit is inversely proportional to $a$.  In Fig.~\ref{fig.durations}, a peak in the distribution of user mark durations can be seen at about six days.  However, the geometrical probability of observing the transit of a planet with the corresponding semi-major axis is extremely small, on the order of $10^{-6}$.  Thus, we expect all or nearly all of the markings with such long durations to be identification mistakes.  Visual inspection of a subset of these marks supports the assumption that most, if not all, correspond to cases in which users have either brought attention to a long data gap or mistaken stellar activity dips (such as those due to starspots) for exoplanet transit signals.  These markings are filtered out using an adjustable ``duration cutoff'' hyperparameter, for which we found the optimal value was about $2.5$ days (see Appendix~\ref{sec.optimize}).  This duration cutoff may not be applicable to datasets from other projects, or a minimum duration cutoff may be more contextually applicable.  This can be changed easily, if desired.

Similarly unique to the PH dataset, each 30-day subject is a fragment of a longer light curve with a designated quarter, and the light curve itself from a designated \kepler target.  The quarter and \kepler ID corresponding to each metafeature are thus recorded in the metafeature list for post-pipeline reference, but may not have any analogous function for other datasets.

The general methodology for consolidating user marks into metafeatures is as follows.  Any user mark that is not filtered out by either the duration cutoff or data gap check is assumed to indicate the presence of a feature that should be scored.  If only one user marked the feature, the mark is recorded exactly as is.  Where multiple user marks overlap on a subject, they are assumed to indicate the same feature if the midpoints of each mark differ by no more than a specified ``midpoint tolerance.''  For a given user mark, we find all marks made by other users that satisfy the midpoint criterion, forming a ``match list'' for the given mark.  Then, for each mark in the match list, we obtain new match lists.  Marks with identical match lists are consolidated to construct the resulting metafeature, by averaging the bounds of all marks in the list.  No user mark contributes to more than one metafeature.  A user mark that matches more than one separate feature is included in the consolidation of neither, comprising its own metafeature, to prevent smearing separate features closer together.  Finally, each metafeature is assigned a unique feature ID.

Users often mark non-planetary-like features, including data gaps, as mentioned above, as well as transits from EBs and brightness variations associated with intrinsic stellar variability.  Though we do attempt to filter some of these marks out of the metafeature list, it is not strictly necessary; a well-constructed scoring process ensures that, if only a small number of users with low weights have marked them, those metafeatures will be dispensed from the pipeline with a very low score.

\begin{figure*}
\centering
\includegraphics[width=1.0\linewidth]{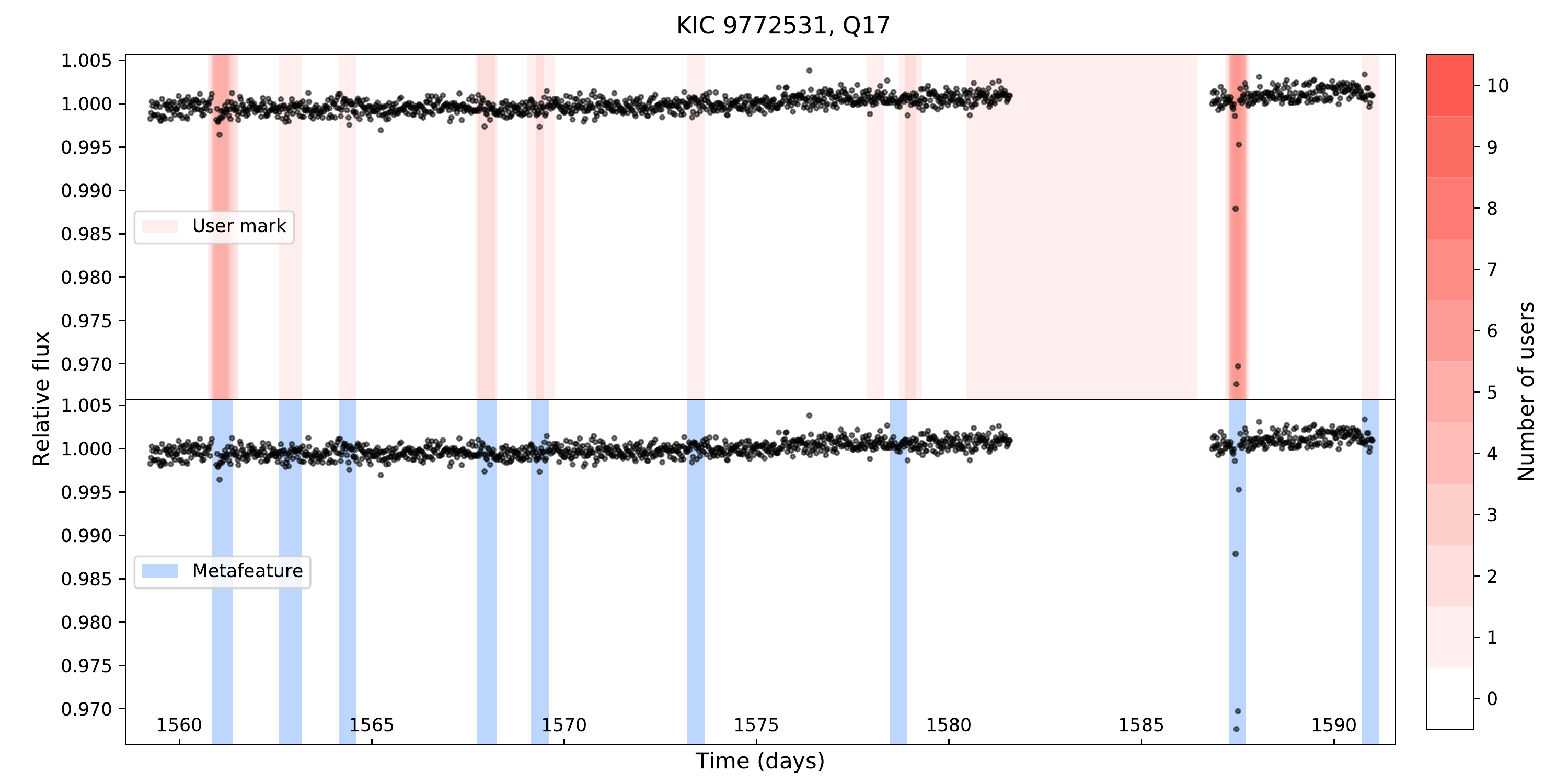}
\caption{\kepler relative flux measurements of KIC~9772531 during Quarter~17, overlaid with all user markings (top) and with those markings consolidated into metafeatures to be scored (bottom).  Hyperparameters for this set of metafeatures were midpoint tolerance $=0.8$~days and duration cutoff $=2.5$~days.\label{fig.usermarks}}
\end{figure*}

Two hyperparameters are adjustable in the construction of the list, as defined above, with the following default values:
\begin{itemize}
    \item Midpoint tolerance:~~1.2 days.
    \item Duration cutoff:~~1.0 day.
\end{itemize}
The default values, which were used to produce the results discussed throughout this work, were chosen based on the optimization study described in Appendix~\ref{sec.optimize}.  The resulting metafeature list, for the PH dataset, contains 486,504 unique features.  Figure~\ref{fig.usermarks} shows the PH user marks on the \kepler Quarter~17 light curve for KIC~9772531, to illustrate how user marks are culled and consolidated into metafeatures for scoring.

\subsection{User weighting}\label{sec.weighting}

The next step in the process is calculating the user weights that will seed the first calculation of metafeature scores, based on matches to the synthetic transit signals in simulations.  It's important to note, at this stage, that these weights can be calculated \textit{only} for those users who did see at least one simulation.  Due to the random nature of when users were shown simulations, a user who visited the website and classified only one or two light curves before leaving may or may not have seen one.  Users who have never classified a simulation comprise about 54\% of the total M-dwarf user base, and these users generated about 23\% of the total M-dwarf classifications (see Fig.~\ref{fig.classsynthvsreal}).  Weights for such users are calculated iteratively later in the pipeline, by the consensus agreement method described in Section~\ref{sec.iteration}.

The general strategy for calculating user weights from simulations, adapted from S12, is as follows:  All users begin with a weight of 1.0.  A user's weight is increased if they make a mark that matches a synthetic transit, detailed in Section~\ref{sec.upweighting}.  A decay function, described in Sec.~\ref{sec.upweighting} and Appendix~\ref{sec.optimize}, adjusts the relative increase in weight for multiple correctly marked synthetic transits in the same light curve.  The user's weight is decreased if they mark spurious features, detailed in Section~\ref{sec.downweighting}.  The user's weight is unaffected if they mark nothing at all on a simulation.  All user weights are then normalized by the mean of the weight distribution, as described in Section~\ref{sec.norm}.  These processes are all contained in the pipeline piece \texttt{user-weighting.py}

\subsubsection{Upweighting}\label{sec.upweighting}

A simulation light curve may contain multiple synthetic transit signals, and each user may make multiple user marks on a simulation.  For an individual user, marks that match synthetic signals are identified by two criteria:
\begin{itemize}
    \item The midpoint of the mark must lie within the specified tolerance of the midpoint (1 day, by default) of the synthetic signal;
    \item The mark must have non-zero overlap with the duration of the signal.
\end{itemize}
In the event that a user has made more than one mark that satisfies these conditions, the one with the closest midpoint to that of the synthetic signal is designated as the match, provided that it has not already been designated a match to another signal.  No user mark is designated a match for more than one synthetic signal.  This process is repeated for all synthetic signals in the simulation.

The user's weight is increased for marks that match synthetic signals.  For each signal, a quantity called signal ID completeness (SIDC) is calculated, defined as the ratio of the number of users who correctly identified the signal to the number of users who saw the simulation containing it.  If a small fraction of users identified the signal correctly, then the SIDC for that signal is small; if the signal was identified correctly by most users, then the SIDC is near unity.  The ``raw'' upweight for correctly identifying a synthetic signal is given by\footnote{Unity is subtracted to account for situations in which all users correctly identify a synthetic signal (such that SIDC $=1$), and thus none of the users who saw the signal can be identified as more skilled than any other.}
\begin{equation}\label{eq.raw-upweight}
    \text{raw upweight} = \frac{1.0}{\text{SIDC}} - 1.
\end{equation}
After two signals have been found in a light curve, it is reasonable to assume that the user knows where to look for others due to the periodic nature of exoplanet transits.  Occasionally, users will even stop marking additional signals after they've found a sufficiently large number of them in a subject, seemingly content to have noted the presence of interesting features in general.  Thus, allowing large, compounded increases in user weight is not necessarily an accurate representation of a user's skill.  To compensate for this, all of the user's raw upweights from a given simulation are sorted from largest to smallest, and then each upweight is scaled by a decay function.  That is,
\begin{equation}\label{eq.upweight}
    \text{upweight} = (\text{raw upweight})\cdot f_\text{decay}(k),
\end{equation}
where $k$ is the ordered index of the upweight in the sorted list, starting at $k=0$.  The form of the decay function is
\begin{equation}\label{eq.decay}
    f_\text{decay}(k) = \begin{cases}
        g(k) &\quad\text{if~}k \le 3, \\
        g(3)~e^{-1.5(k-3)} &\quad\text{if~}k>3,\\
        \end{cases}
\end{equation}
where
\begin{equation}\label{eq.decay-inner}
g(k) = 1.03 - 0.03~e^{1.07~k}.
\end{equation}
The highest upweight is given to the user in its entirety, the second highest is given at about 94\% of its raw value, the third is given at about 78\%, and the fourth at about 29\%.  Any further synthetic signals identified in the light curve decrease with an inverse exponential scaling.  

The coefficients in Equation~\ref{eq.decay-inner} were chosen through author trial and error.  Appendix~\ref{sec.decay} discusses details of this decay function, and compares user weight seed distribution results obtained using a variation with a cutoff, and no decay function whatsoever.  We note that, for projects seeking to identify aperiodic or irregular signals, simply setting the decay function to unity will allow all raw upweights to be applied, unaltered, to users' weights.

\subsubsection{Downweighting}\label{sec.downweighting}

A user's weight is not decreased if the user fails to find a synthetic signal.  The normalization process following the upweighting and downweighting effectively lowers the weights of users who have made no markings, so to explicitly downweight for failure to identify signals is to punish users redundantly.  Instead, downweights are given only for incorrect markings on synthetic lightcurves, which indicate spurious features.  For each marking a user made that does not correspond to a synthetic signal, the magnitude of the user's downweight is increased by 1.

Note that this methodology differs considerably from that of the PH {\it TESS} pipeline, in which both incorrect markings and failure to find synthetic signals affect a user's weight explicitly.  We consider the method of \citet{bib.eisn2020b} a viable alternative strategy, and an interesting opportunity for potential future exploration.

\subsubsection{Normalization}\label{sec.norm}

After upweights and downweights have been calculated for all users who saw synthetics, both sets of numbers are normalized so that the mean of each distribution is 1.  That is, if $n$ is the number of users who have seen synthetics, then the normalized upweight for user $i$ is
\begin{equation}\label{eq.normcombined}
    (\text{norm. upweight})_i = \frac{(\text{upweight})_i}{\frac{\sum_n (\text{upweight})_n}{n}}~,
\end{equation}
and likewise for the user's downweight.  The ratio of the user's upweight and downweight form their ``combined'' weight, and finally, the distribution of combined weights is normalized as well.

For the PH users in this sample, the weight of a user who saw simulations but received neither any upweights nor downweights from their classifications is $\approx 0.7874$ after normalization.  This ``base weight'' is used as the dummy weight for users who have not seen synthetics in the first round of metafeature scoring.  At this stage, the user weight table is also refined to include lists of which metafeatures were classified by each individual user.  This saves a significant amount of time when the weighting is coupled to the scoring.

\subsection{Metafeature scoring}\label{sec.scoring}

As in S12, scores for each feature, including both synthetic and non-synthetic signals, are calculated using the basic formula

\begin{multline}\label{eq.score}
    s_{k,j} = \frac{1}{W_k} \sum_i w_i \cdot j*, ~~~~~~~~~~ j* = 
        \begin{cases}
            0 & \text{if $j_i \neq j$} \\
            1 & \text{if $j_i = j$} \\
        \end{cases}
        \\\text{where $i$ = all users who viewed the subject containing $k$.}
\end{multline}

Like S12, a vote $j$ score is calculated for each metafeature $k$, where $j_i$ ~$=$~``yes'' or ``no'' indicates that the metafeature either was or was not marked by user $i$,
respectively, $w_i$ is the weight of user $i$, and $W_k$ is defined as the sum of user weights for all the users who classified the subject containing metafeature $k$.  A user who classified the subject but did not mark the metafeature in question has effectively voted ``no'' for that metafeature.

All users contribute to this scoring process, which is done by the pipeline piece called \texttt{score-seeds.py}.  As the score for each metafeature is calculated, the users who have classified the light curve containing that metafeature are recorded in the scoring dataframe, divided into a \texttt{usersyes} group (those who identified the feature) and a \texttt{usersno} group (those who did not) for easy access later.  Much like the recording of the IDs of metafeatures classified by each user, this saves time in the coupled process.

\subsection{Coupled scoring-weighting process}\label{sec.iteration}

The initial scores calculated for each metafeature act as seeds, which are used to further adjust the weight of each user based on the user's agreement with majority votes.  This adjustment is mathematically adapted from S12.  For user $i$ with weight $w_i$, a new weight is recalculated:
\begin{multline}\label{eq.weight-adj}
    w_{i,\text{new}} = \frac{A}{N_i} \sum_k s_k(j \text{ for metafeature $k$ chosen by user $i$}), \\ \text{where $k$ = all metafeatures reviewed by user $i$},
\end{multline}
where $j$ is the vote that user $i$ cast for metafeature $k$.  That is, if a user voted ``yes'' for metafeature $k$, the ``yes'' score is added to the sum; if the user voted ``no'' for the metafeature, it is the ``no'' score for that feature that is included in the sum instead.  The normalization factor $A$ is defined to force the mean of the weight distribution back to unity.  Unlike S12, $N_i$ is defined as the number of metafeatures classified by user $i$, rather than the number of light curves.

In practice, the normalization factor $A$ is applied after all user weights have been adjusted for vote agreement.  The entire distribution is then renormalized, and the metafeature scores recalculated based on the new user weights, using the same method as in Eq.~\ref{eq.score}.

\begin{figure}
\centering
\includegraphics[width=1.0\columnwidth]{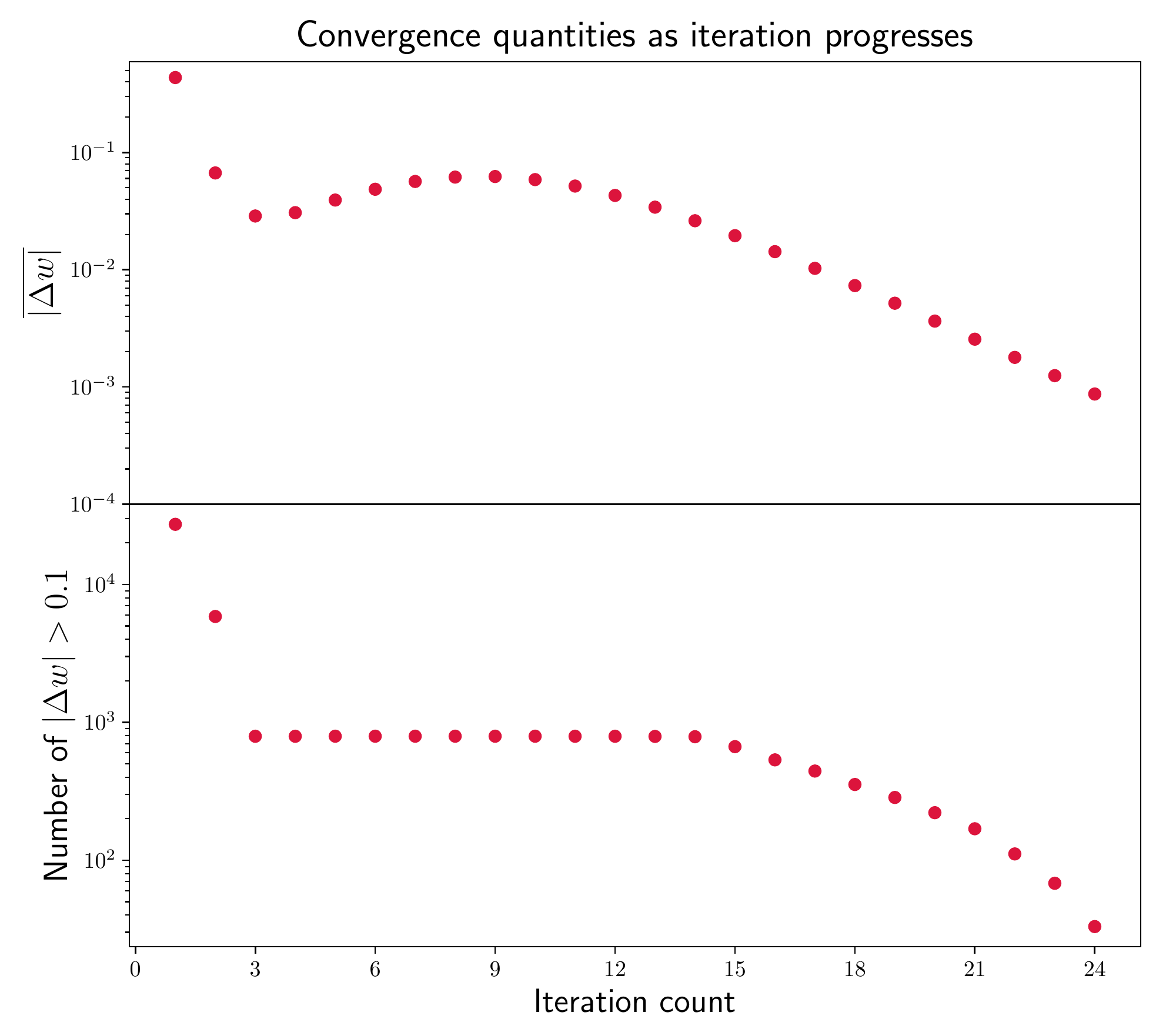}
\caption{Evolution of convergence quantities over the course of the coupled weighting-scoring process.  Top:  The mean magnitude of the user weight differences from the first iteration up to convergence, when the mean magnitude difference reaches $<10^{-3}$.  Bottom:  The number of users whose weights are still changing by more than $0.1$ units after each iteration.\label{fig.meanmagdiffs}}
\end{figure}

\begin{figure*}
\centering
\includegraphics[width=1.4\columnwidth]{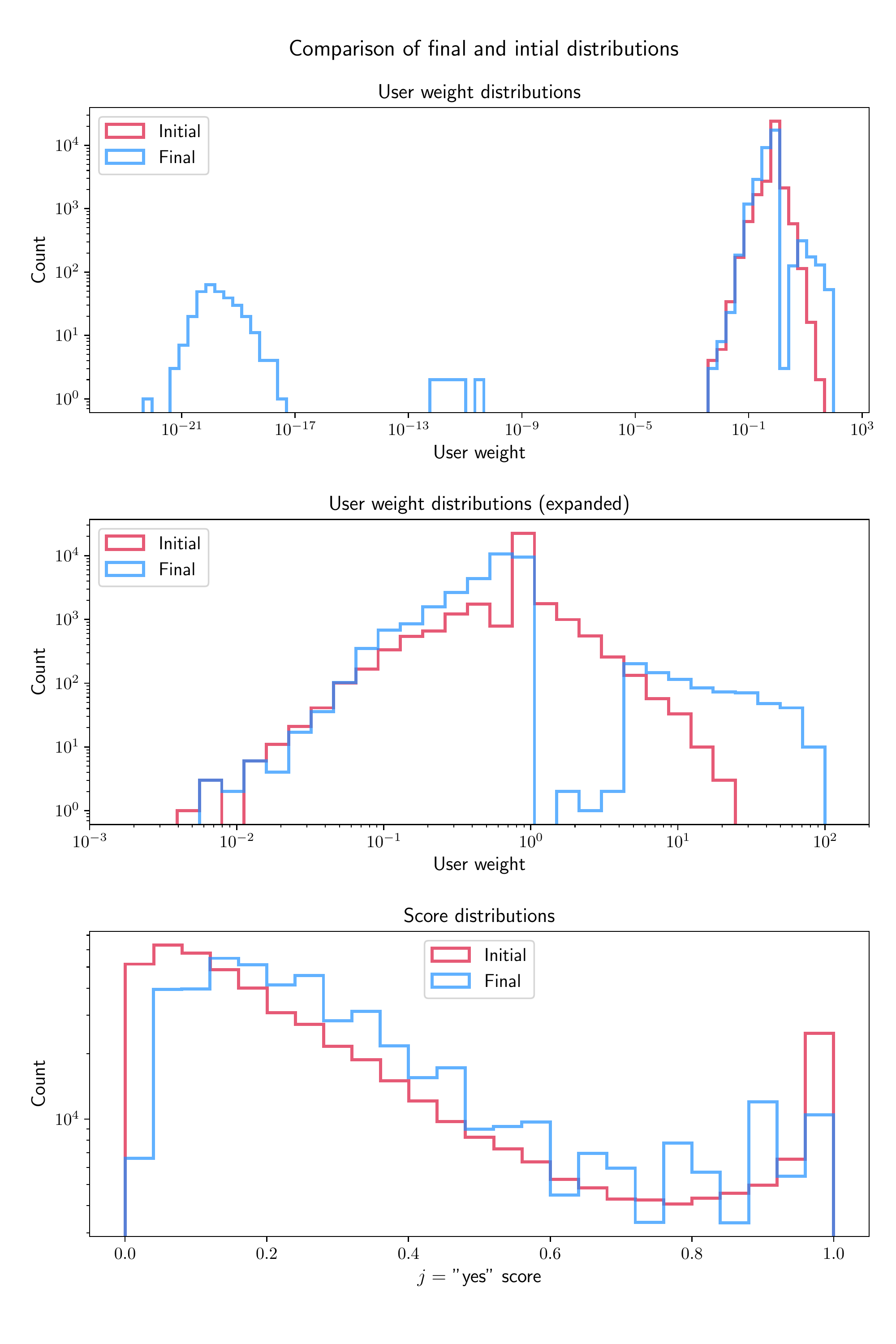}
\caption{Comparison of user weight (top) and metafeature score (bottom) distributions at the beginning and end of the coupled iteration.  The highest subset of user weights is expanded in the middle panel.  Both the user weights and the metafeature scores undergo the most dramatic changes during the first few iterations.  Over subsequent iterations, subsets of users break away from the main distribution of weights (one of which approaches zero, and the other of which increases liberally, comprised of users who classified no real data), while the weights of the vast majority change negligibly.\label{fig.init-v-final}}
\end{figure*}

The process is repeated until the mean absolute difference in weights between iterations is less than $10^{-3}$, at which point the weights are considered to have converged.  For the PH classifications, convergence was achieved after 24 iterations.  
Figure~\ref{fig.meanmagdiffs} shows the mean of the distribution of absolute weight differences as a function of iteration number.  The large, early drop corresponds to the most significant change in the distributions of both user weights and metafeature scores.  Over these first iterations, the weight distribution is narrowed from a range between roughly 0.01 and 22 to a range of about 0.01 to 1.5.  
Between iterations 3 and 15, the majority of user weights change negligibly (bottom panel of Fig.~\ref{fig.meanmagdiffs}). During this period, the weights of a small number of users become arbitrarily closer to zero, and the weights of another small subset of users (who have classified only simulations, no real data) increase liberally.  The latter continues until near the last few iterations.  These two subsets of users are visible in blue in the top pane of Figure~\ref{fig.init-v-final}, which shows how the distributions of weights and scores change from initialization to convergence of the coupled weighting-scoring process.\footnote{Animations of the changes in PH user weight distribution, weight difference distribution, and score distribution over the course of the entire process are publicly available on the PH pipeline GitHub.}

The consensus agreement weighting method implemented in this coupled process dominates the final weight and score distributions, the latter to a stronger degree than the former.  We performed a test run of SATCHEL using only weights from simulations (with uniform dummy weights for users who saw no simulations) and another test run using entirely uniform dummy weights as seeds for the consensus agreement weighting process, and compared the resulting weight and score distributions to each other and to the final distributions when both methods are used.  We found that, when users are weighted based only on simulations, SATCHEL recovers slightly more synthetic signals (as discussed below in Sec.~\ref{sec.recovery}), while the consensus-only run recovered slightly more KOI signals.  Whether one or the other is used, or both, the differences in recovery are well within Poisson errors.

Converging on only the mean stability of the user weight distribution, rather than the stability of all the user weights individually, allows for the possibility that a coincidental pseudo-stability may be reached, in which one or more weights are still increasing by a large amount, but other weights are decreasing by a commensurate amount, balancing out the distribution as a whole.  One of the outputs of each iteration is a count of how many user weights are still undergoing ``large'' changes.  In the case of the PH data, changes of more than about 0.1 units are considered large; this threshold may be different for other data sets.  Whether or not such large changes are still occurring, the iteration is currently built to accept the convergence.  For the PH data, convergence was achieved while the number of users with weights still changing by more than 0.1 units was 33.  Those weights belonged to users who had classified only simulations (typically only one subject), and no real data, so the scores of all the real metafeatures were already stable, affected negligibly by such changes.

Another circumstance which may potentially arise is that of a resonance loop between the weights and scores, in which the weight adjustment causes a change in metafeature scores that exactly reverts the weights to their prior state after the next adjustment, and so forth.  The mean of the weight distribution is compared to that of the iteration before last (as soon as the number of iterations elapsed is high enough to do so), and the same is done for the mean of the score distribution.  If both means are the same for three alternating past iterations within a tolerance of $10^{-3}$, a set of small, random perturbations are added to the metafeature scores to attempt to break the resonance loop, and this attempt is noted in the terminal output.  For the PH data, a resonance loop did not arise, so random perturbations were not necessary.

Once convergence is reached, the final states of both the user weights and resulting metafeature scores are saved, along with several plots and animations that may be useful.  If the loop is exited prior to convergence, plots and animations constructed using weight and score data from iterations up to that point are saved, along with the state of the weights and scores from the last complete iteration.  A log file containing potentially significant quantities for each iteration (including, among other things, the mean magnitude weight difference, the means of the weight and score distributions, and the total time elapsed) is also saved as output.

\section{Measuring performance}\label{sec.recovery}

In the following subsections, recovery is assessed against both the simulations (Sec.~\ref{sec.simreco}) and against the known population of KOIs (Sec.~\ref{sec.KOIreco}).  Both types of light curve subjects were displayed the same way in the PH interface, to minimize potential user biases in classification, and other than the use of simulations in calculating the user weight seeds, both types of subjects were treated identically throughout the rest of the pipeline.  A known signal was considered recovered if a metafeature that overlapped the signal by at least 50\% had a final score of 0.5 or greater.  This score cutoff was chosen to balance maximum recovery of known signals with minimum false positive contamination (see Appendix~\ref{sec.scorecutoff} for details).  Out of 486,504 metafeatures, 88,171 met this requirement, 3522 of which correspond to transits of known KOIs and 13,616 of which correspond to synthetic transit signals in simulations.

\begin{figure}
\centering
\includegraphics[width=1.0\columnwidth]{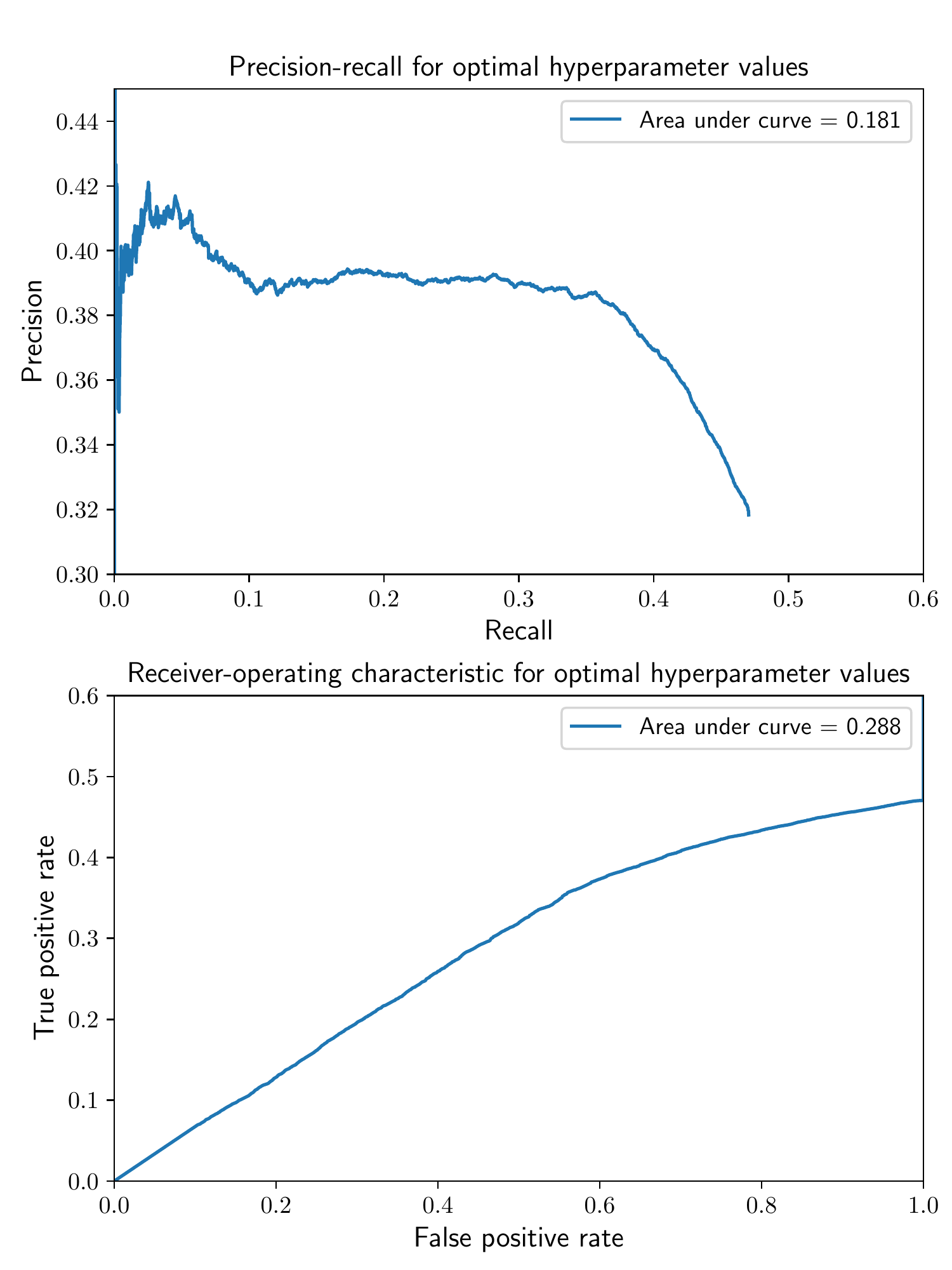}
\caption{The Receiver-Operating Characteristic (ROC) and Precision-Recall (PR) curves for the optimized SATCHEL pipeline.  The area under each curve was calculated using the trapezoidal rule.\label{fig.ROC-PR}}
\end{figure}

\begin{table}
	\centering
	\caption{Performance statistics for the SATCHEL pipeline, with optimal hyperparameters (duration cutoff = 1.0 day, midpoint tolerance = 1.2 days).}
	\label{tab.stats}
	\setlength\extrarowheight{2pt}
	\begin{tabular}{|c|c|} 
		\hline
		Statistic & Value \\
		\hline
		Area under ROC curve & 0.288 \\
		Area under PR curve & 0.181 \\
		Average precision score & 0.445 \\
		\hline
	\end{tabular}
\end{table}

Taking both synthetic signals from simulations and signals from KOIs as ``ground-truths,'' we quantify the performance of the pipeline as a whole through inspection of the Receiver Operating Characteristic (ROC) curve and the Precision/Recall (PR) curve, using the \texttt{sklearn} Python package.  We consider the score threshold as a binary classifier, such that a metafeature in a simulation subject or KOI subject assigned a score higher than the threshold is defined as a ``positive.''  Such a metafeature that matches a ground-truth signal is then a true positive, while one that does not is a false positive.  A false negative may then be defined as a ground-truth signal that was either not identified as a metafeature or a matching metafeature which was assigned by the pipeline a score lower than the threshold value.  ``True negatives'' would, in principle, be features in subjects that were either not identified as metafeatures or non-matching metafeatures that were assigned scores lower than the threshold value.  As the number of unmarked features is not quantifiable, neither is the number of true negatives.

The ROC curve is constructed by plotting the fraction of true positives out of the positives vs. the fraction of false positives out of the negatives, at various threshold values.  The precision is, effectively, the pipeline's ability to correctly sort metafeatures into positives and negatives.  The recall is the pipeline's ability to correctly identify {\it all} ground-truth signals.  For this binary classifier, the ROC and PR curves are shown in Figure~\ref{fig.ROC-PR}.  The area underneath each curve (AUC) can be used as a single-number metric to gauge pipeline performance.  The closer the AUC is to unity, the better the performance.  The ROC AUC for SATCHEL is 0.288; the PR AUC is 0.181.  We also calculate the average precision score, for ease of performance comparison to future pipelines.  A summary of these values and the hyperparameters that optimize them is given in Table~\ref{tab.stats}.

\begin{figure}
\centering
\includegraphics[width=1.0\columnwidth]{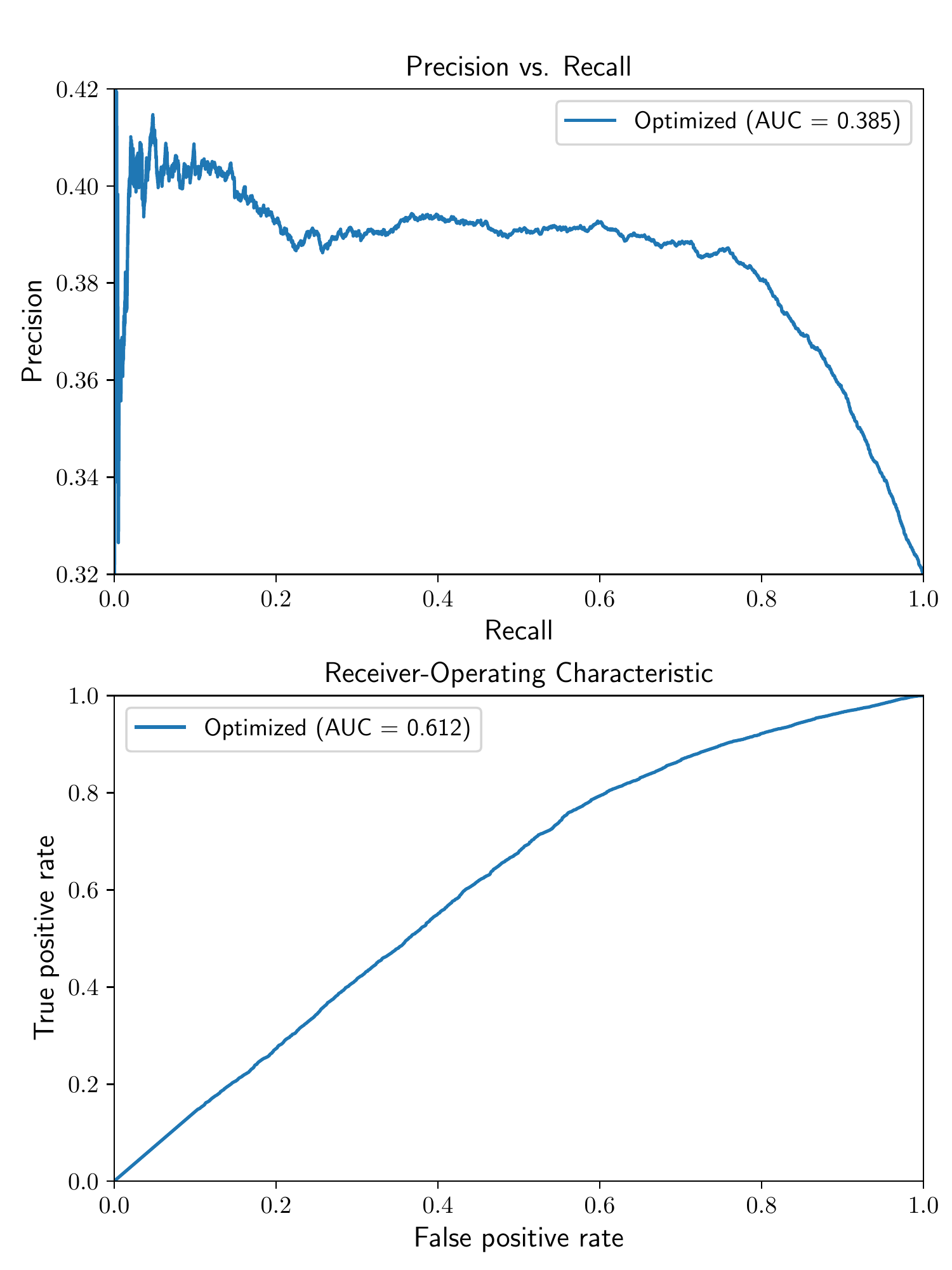}
\caption{The Receiver-Operating Characteristic (ROC) and Precision-Recall (PR) curves for the optimized SATCHEL pipeline, not including transit signals that were not found by PH volunteers.  The area under each curve was calculated using the trapezoidal rule.\label{fig.ROC-PR-foundonly}}
\end{figure}

Generally, the ROC curve for a classifier passes through the points $(0,0)$ and $(1,1)$, indicating that, with the strictest threshold, no false positives are recovered but neither are any true positives, and with most lenient threshold, all true positives are recovered, as well as all false positives.  For the SATCHEL pipeline, this cannot be achieved, because there is no threshold for which all simulations and KOI transits were found by PH volunteers.  If unfound transit signals are removed from the analysis, the ROC curve behaves in the standard way, as shown in Figure~\ref{fig.ROC-PR-foundonly}.

S12 found that, rather than exoplanet radius, the signal depth $\delta = R_p^2/R_\star^2$ (where $R_p$ is the radius of the planet and $R_\star$ is the radius of the host star) is the true observable for PH volunteers.  Though our sample is comprised of only M-dwarf stars, the range of stellar radii is still significant, spanning $0.13$ to $0.78~R_\odot$ \citep{bib.gaid2016,bib.berg2020}.   Furthermore, M-dwarf light curves can be afflicted with noise from sources such as sunspots, flares, and other magnetic activity to an extent that is generally more extreme than stars of other spectral types \citep{bib.niel2013}.  Signal-to-Noise Ratio (SNR) may, therefore, be even more directly related to the identifiability of transit signals by PH volunteers.  In the below subsections, we consider SNR {\it for each individual transit only}, calculated as the ratio of the signal depth $\delta$ to the Combined Differential Photometric Precision \citep[CDPP,][]{bib.koch2010,bib.chri2012}, omitting the factor of $\sqrt{n_{tr}}$ that is commonly included in analysis of \kepler light curves to account for the strengthening of the phase-folded signal containing a number $n_{tr}$ of transit signals over the duration of observation \citep{bib.howa2010,bib.chri2012}.  The PH interface did not offer such algorithmic enhancement of signals, and thus we would not expect such a factor to be applicable in any potential detection trends.  It is likely that the presence of multiple signals in a single subject increases the detectability of individual signals, as discussed in Sec.~\ref{sec.upweighting}, but quantifying this effect may be either impossible or prohibitively difficult, as behavior varies greatly from one volunteer to another.  If this bias toward short-period signals exists, its effect on pipeline completeness is limited only to simulations and KOIs with periods shorter than about 15 days, due to the 30-day time baseline of PH subjects.

\subsection{Simulation recovery}\label{sec.simreco}

\begin{table*}
	\centering
	\caption{Recovery frequency of synthetic exoplanet transit signals.  Note that the binning here is not identical to that of Fig.~\ref{fig.recotiles-simsignals}}
	\label{tab.simfreq}
	\setlength\extrarowheight{2pt}
	\begin{tabular}{lccccc} 
		\hline \hline
		Planet Radii & \multicolumn{5}{c}{Orbital Period (days)} \\
		\cline{2-6}
		$(R_\oplus)$ & $0.7 \leq P<200$ & $200 \leq P<400$ & $400 \leq P<600$ & $600 \leq P<800$ & $800 \leq P<1000$ \\
		\hline
		$1 \leq R<2$ & 848/2408 & 324/883 & 289/594 & 368/666 & 274/523 \\
		$2 \leq R<3$ & 2221/3160 & 1015/1272 & 590/689 & 564/664 & 491/576 \\
		$3 \leq R<4$ & 668/764 & 127/142 & 95/116 & 169/194 & 131/141 \\
		$4 \leq R<5$ & 717/748 & 359/378 & 110/124 & 135/148 & 88/88 \\
		$5 \leq R<6$ & 1078/1133 & 324/326 & 208/228 & 135/135 & 172/172 \\
		$6 \leq R<7$ & 216/231 & 55/57 & 39/39 & 22/22 & 26/26 \\
		$7 \leq R<10$ & 466/488 & 220/235 & 89/108 & 87/94 & 81/81 \\
		$10 \leq R<15$ & 496/516 & 97/97 & 124/124 & 28/28 & 70/71 \\
		\hline
	\end{tabular}
\end{table*}

As mentioned in Sec.~\ref{sec.interface}, the synthetic signals that populate the simulations were constructed from a range of exoplanet radius and orbital period values.  Radius values were pulled at random from a distribution between 0.5~\re\ $<R_p< 12~$\re\, and orbital periods were pulled from a distribution between $0.7~\text{days} < P < 1000~\text{days}$.\footnote{The distributions were not uniform, but the authors have been unable to obtain clarification on the exact binning.}  Neither eccentricity nor orbital inclination were varied; orbits were circular only, and transits passed perfectly along the diameter of the star.  For a given synthetic exoplanet, the orbital phase was drawn randomly from the range $0$ to $2\pi$ until at least one transit occurred within the 30 day window of the simulation subject.  The duration and flux loss of the transit signal were calculated based on the orbital period, exoplanet radius, and host star radius, and the flux loss was subtracted directly from the PDCSAP flux values of the stellar light curve.  No artificial noise was added.  The number of synthetic signals corresponding to exoplanet radii and periods falling in a range of bins is listed in Table~\ref{tab.simfreq}.

\begin{figure}
    \includegraphics[width=1.0\columnwidth]{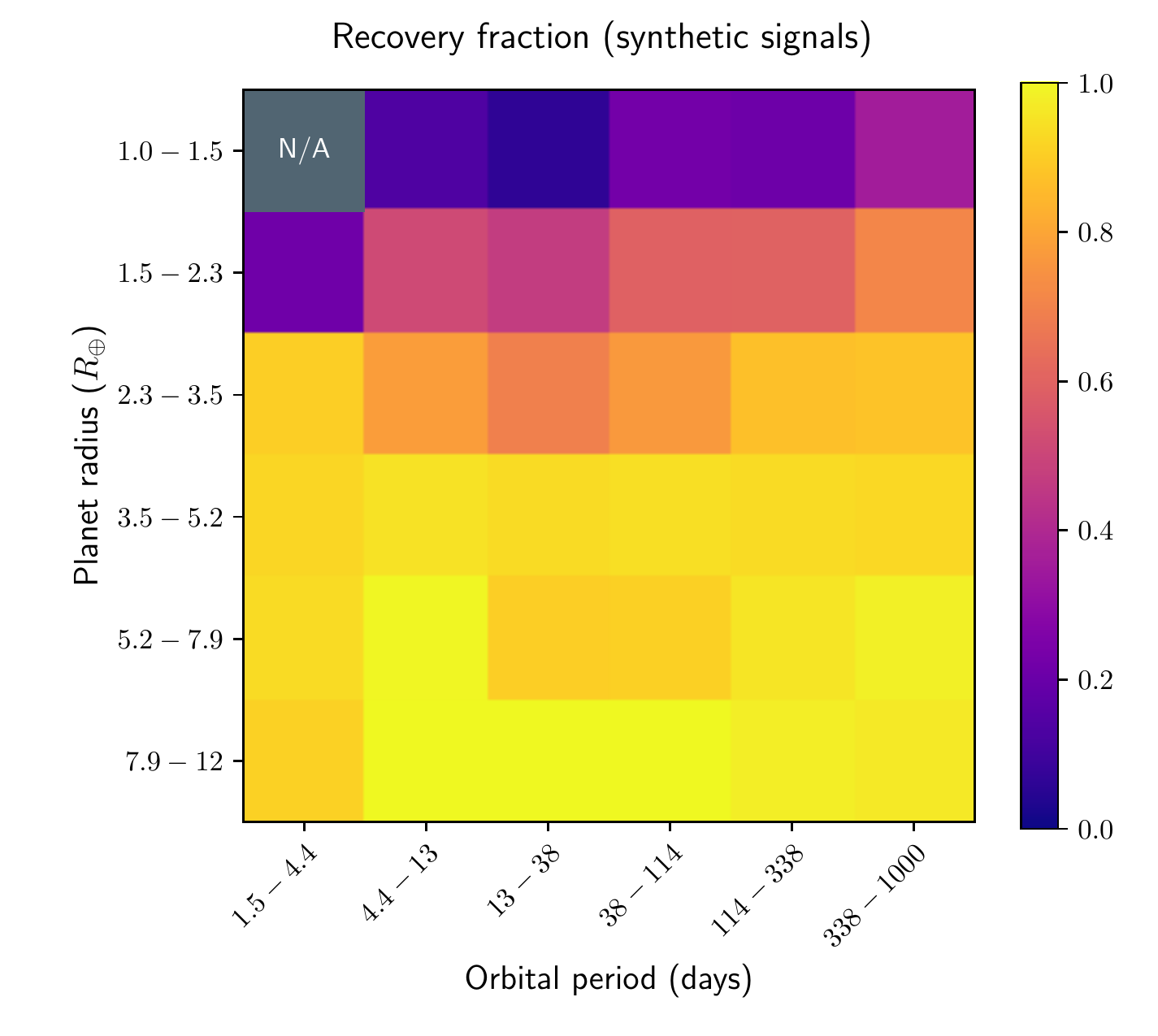}
    \caption{Recovery of synthetic signals as a function of both exoplanet radius and orbital period, in which a signal is considered recovered if a matching metafeature was assigned a score of 0.5 or higher by the PH pipeline.  Grey bins labeled ``N/A'' correspond to bins in which there were no synthetic signals to be recovered.  Binning in both radius and period are logarithmically even.\label{fig.recotiles-simsignals}}
\end{figure}

\begin{figure*}
    \centering
    \includegraphics[width=1.0\linewidth]{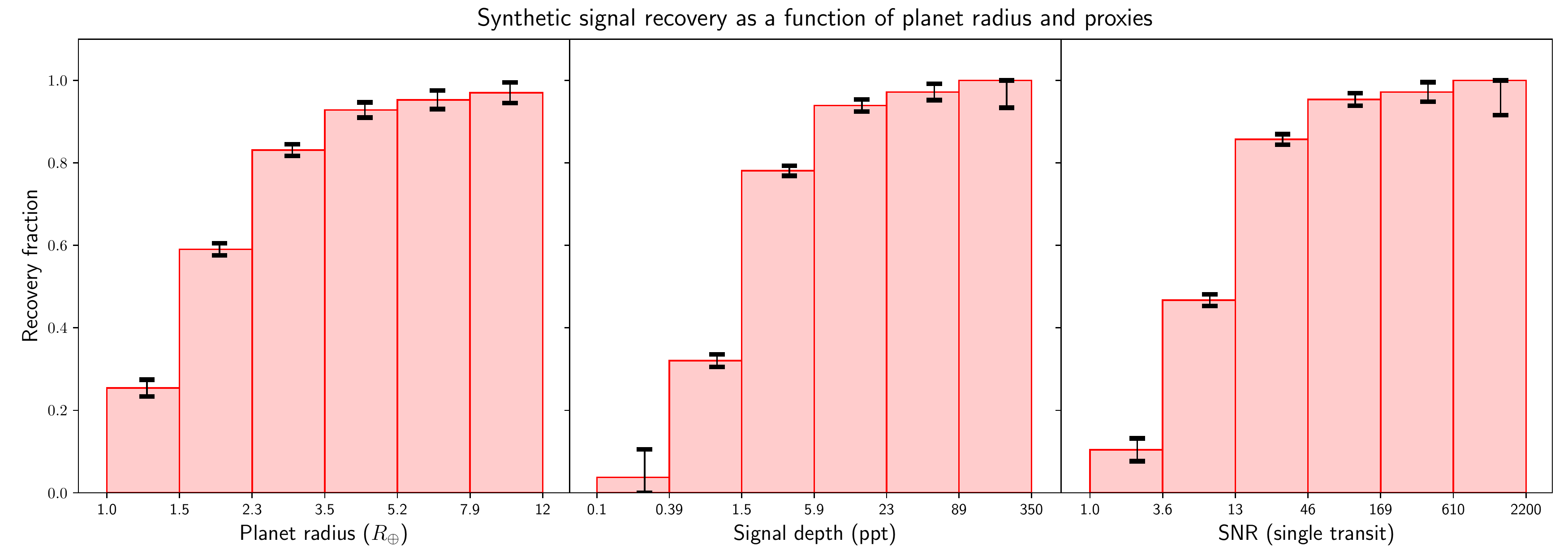}
    \caption{Recovery of synthetic signals as a function of radius (left, in $R_\oplus$), depth (middle, in parts-per-thousand), and SNR (right), all marginalized over orbital period by sum, with Poisson errors.  SNR is calculated per-signal (see discussion in Sec.~\ref{sec.recovery}).  Binning in radius, depth, and SNR are logarithmically even.\label{fig.recov-simsignals-radproxy}}
\end{figure*}

\begin{figure}
    \includegraphics[width=1.0\columnwidth]{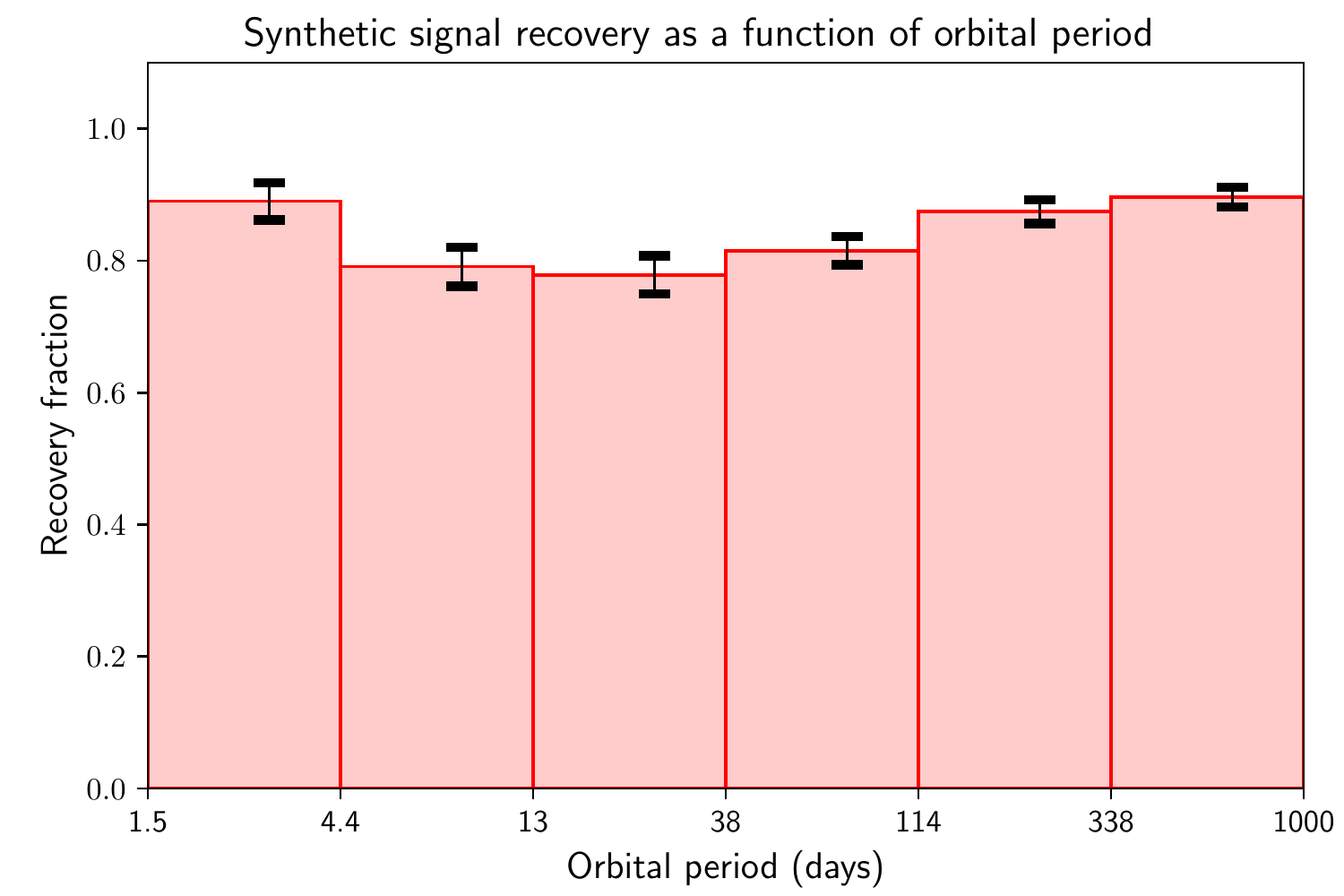}
    \caption{Recovery of synthetic signals as a function of orbital period, marginalized over exoplanet radius by sum.  Binning in orbital period is logarithmically even.\label{fig.recov-period-simsignals}}
\end{figure}

Figure~\ref{fig.recotiles-simsignals} displays recovery rates across the full range of physical properties of the simulations.  Note that the orbital period binning in Fig.~\ref{fig.recotiles-simsignals} is logarithmically even, rather than the linear binning of Table~\ref{tab.simfreq}.  Figure~\ref{fig.recov-simsignals-radproxy} shows the recovery fraction of signals from synthetic exoplanets as a function of radius, transit signal depth, and SNR, all marginalized over orbital period by summing each bin, and with Poisson counting errors shown.  The most prominent trend, clear from all binnings, including that of Table~\ref{tab.simfreq}, is that recovery fraction increases with exoplanet radius.  This is an expected trend; since the signal depth is proportional to the square of the planet radius, larger exoplanets generally produce deeper dips, which are more obvious to the human eye.  Conversely, the small dips from synthetic exoplanets with radii $< 2~R_\oplus$ are easily lost in the noise from typical stellar activity.  The smooth, monotonic evolution of the recovery rate as a function of radius and its proxies, particularly SNR which is arguably the true observable, is exactly what we would expect from visual inspection of the \kepler data.

Above $2~R_\oplus$, the total recovery of all synthetic signals is about 85.8\%.  Above $2.5~R_\oplus$, the total recovery is about 91.0\%.

Figure~\ref{fig.recov-period-simsignals} displays recovery of synthetic signals as a function of orbital period.  There appears to be a trend of decreasing recovery rate with increasing orbital period for short-period signals, with a slight increase or possibly stable rates for signals with orbital periods $\gtrsim 30$~days.  The possible statistical significance of this trend is explored in detail in Paper II.

\subsection{KOI recovery}\label{sec.KOIreco}

The \kepler M-dwarf dataset has been mined extensively for short-period exoplanets \citep[see, e.g.,][]{bib.dres2015,bib.gaid2016,bib.meye2018,bib.hard2019}, and thus, the known population of KOIs is a valuable recovery metric to explore.  We here use the NASA Exoplanet Archive\footnote{\href{https://exoplanetarchive.ipac.caltech.edu/}{https://exoplanetarchive.ipac.caltech.edu/}} (NEA) as a repository for KOI properties such as planetary radius and orbital period, as well as host star radius and spectral type.  We exclude KOIs that have been classified as false positives, and retain both confirmed and candidate exoplanets.  Throughout this discussion, we make no distinction between the latter two.

\begin{figure}
    \includegraphics[width=1.0\columnwidth]{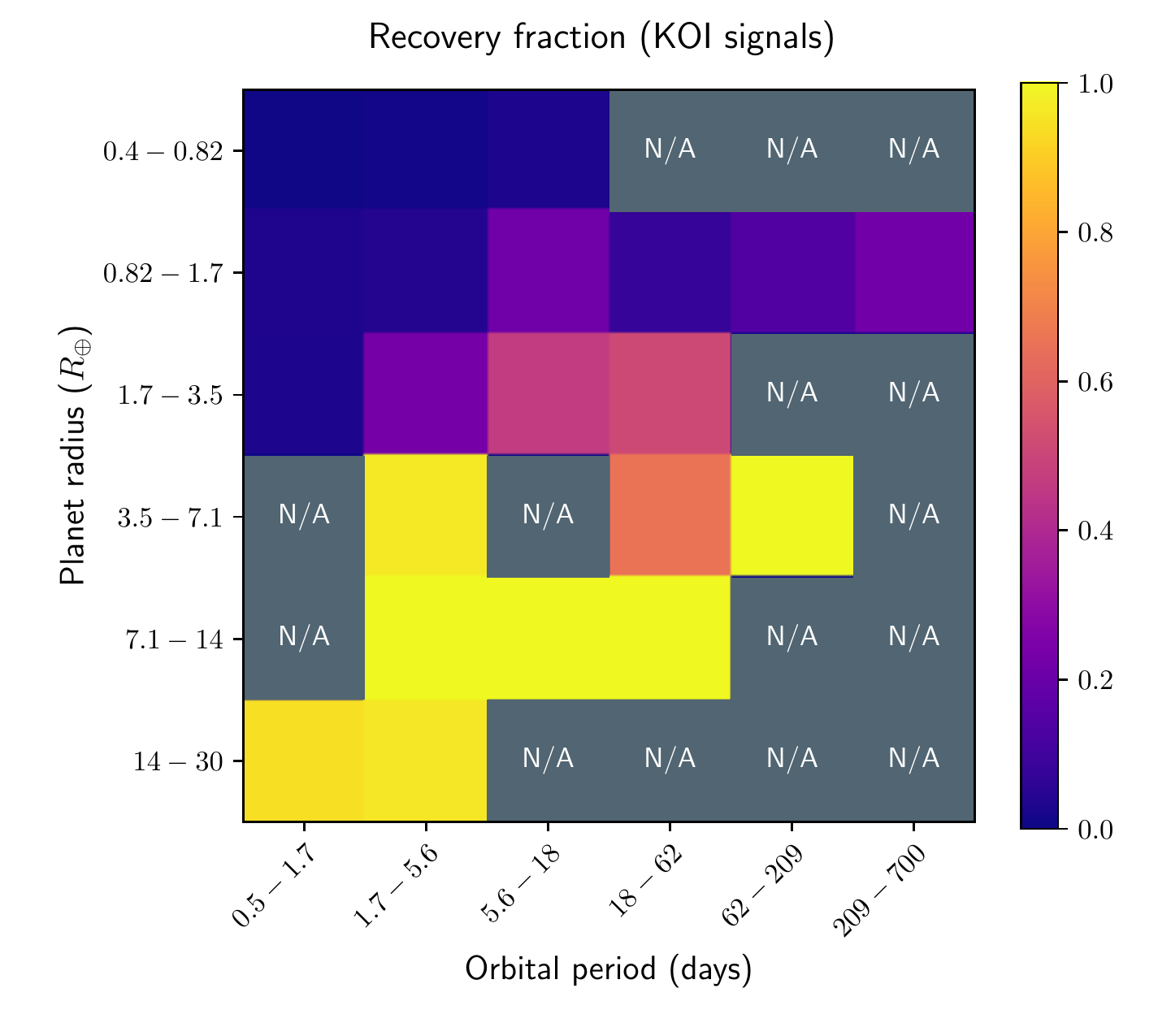}
    \caption{Recovery of synthetic signals as a function of both exoplanet radius and orbital period.  Grey bins labeled ``N/A'' correspond to bins in which there were no KOI signals to be recovered.  Binning in both radius and period are logarithmically even.\label{fig.recotiles-KOIsignals}}
\end{figure}

\begin{figure*}
    \centering
    \includegraphics[width=1.0\linewidth]{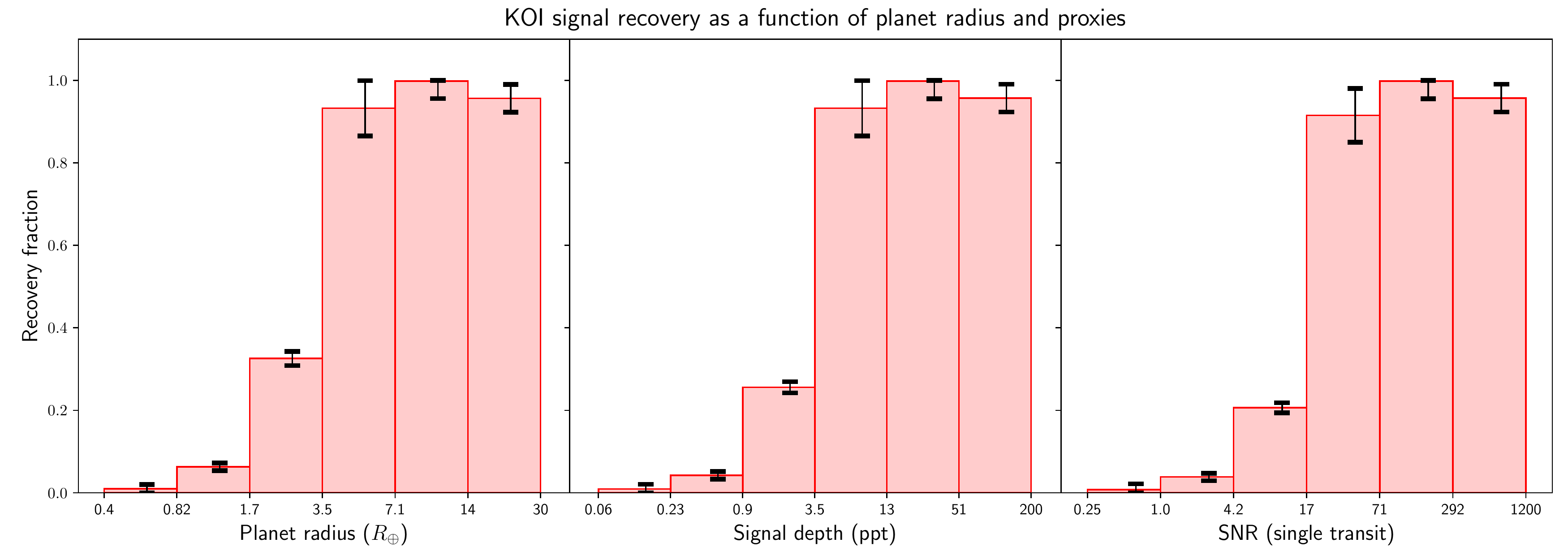}
    \caption{Recovery of KOI signals as a function of radius (left, in $R_\oplus$), depth (middle, in parts-per-thousand), and SNR (right), all marginalized over orbital period by sum, with Poisson errors.  SNR is calculated per-signal (see discussion in Sec.~\ref{sec.recovery}).  Binning in radius, depth, and SNR are logarithmically even.\label{fig.recov-KOIsignals-radproxy}}
\end{figure*}

\begin{figure}
    \includegraphics[width=1.0\columnwidth]{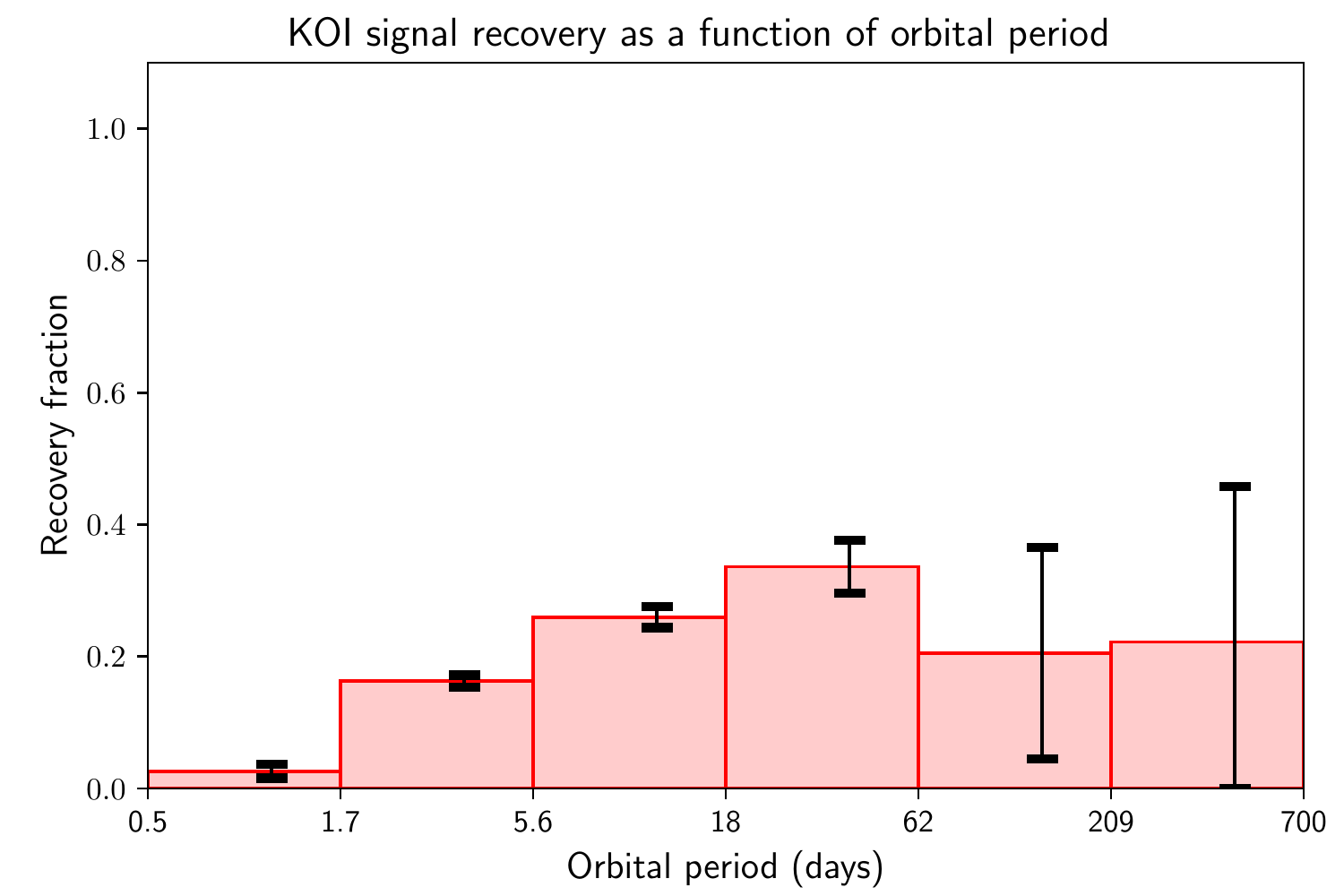}
    \caption{Recovery of synthetic signals as a function of orbital period, marginalized over exoplanet radius by sum.  Binning in orbital period is logarithmically even.\label{fig.recov-period-KOIsignals}}
\end{figure}

There are 112 KOIs listed in the NEA for our stellar sample, with confirmed and candidate exoplanets spanning a range of radii from $0.08<R/R_\oplus<25.0$, and a range of orbital periods from $0.53<P/\text{day}<628$.  A significant fraction of these KOIs reside in a regime of parameter space unrepresented by our simulations, particularly those with very small radii, which makes direct performance and recovery comparisons difficult.  In an effort to retain the maximum number of KOIs for statistical completeness while preserving the comparability of the KOI study to the simulations, the two smallest KOIs with the smallest SNR values (K06705.01 and K04777.01), have been excluded from the following study.  This limits the parameter space to about $0.4<R/R_\oplus<25.0$ in exoplanet radius.  Notably, KOI K06705.01 was recovered by PH volunteers, despite its shallow signals and low SNR (0.045 ppt and about 0.25, respectively).  This is likely owing to its short period, only 1.0 day, which guarantees many transit signals even in a single subject.

Figure~\ref{fig.recotiles-KOIsignals}, like Figure~\ref{fig.recotiles-simsignals}, displays recovery rates across the truncated range of physical properties of the M-dwarf KOIs.  Though the small number of KOIs leaves the plot quite underpopulated, a trend of increasing recovery with planetary radius does seem to be visible.  The trend appears more prominent in Figure~\ref{fig.recov-KOIsignals-radproxy}, which shows the recovery rate as a function of exoplanet radius and its proxies, marginalized over the orbital period.  A corresponding trend of increasing recovery fraction with signal depth and SNR is also visible.  These, again, are trends that we expect, and the similar trend shapes in Figures~\ref{fig.recov-simsignals-radproxy} and~\ref{fig.recov-KOIsignals-radproxy} reinforce our confidence in both the trends themselves and the effectiveness of the simulations as a tool to measure realistic user and pipeline performance, particularly for exoplanets larger than $2~R_\oplus$.  Recovery trends with radius between values of $1-2~R_\oplus$ are less robust in comparison, with significantly smaller recovery rates for KOI signals than synthetic signals, the cause of which remains to be explored in future work.

Recovery of KOI signals as a function of orbital period is shown in Figure~\ref{fig.recov-period-KOIsignals}, with Poisson errors.  Above $2~R_\oplus$, in the radius range that the simulation study suggests PH volunteers can reliably detect, KOIs are quite sparse, and there are none with periods longer than about 77 days.  We are thus unable to make a strong statement about trends in KOI recovery with orbital period, though our results do appear to be loosely consistent with the possibility of a stable recovery rate for signals with periods longer than$~\gtrsim 30$~days.  For KOIs with short periods, however, a trend seems to suggest increasing recovery with orbital period, the reverse of the weak trend seen for short period signals in Fig.~\ref{fig.recov-period-simsignals}.  Whether this is due to the large Poisson uncertainties in the KOI sample or a fundamental difference between the two samples in the physical properties that govern detectability, this discrepancy may be worth exploring in future citizen science studies.

For KOIs larger than $2~R_\oplus$ recovery was 72.1\%, and 94.8\% for KOIs larger than $2.5~R_\oplus$.

\subsection{Completeness}\label{sec.completeness}

\begin{figure}
\centering
\includegraphics[width=1.0\columnwidth]{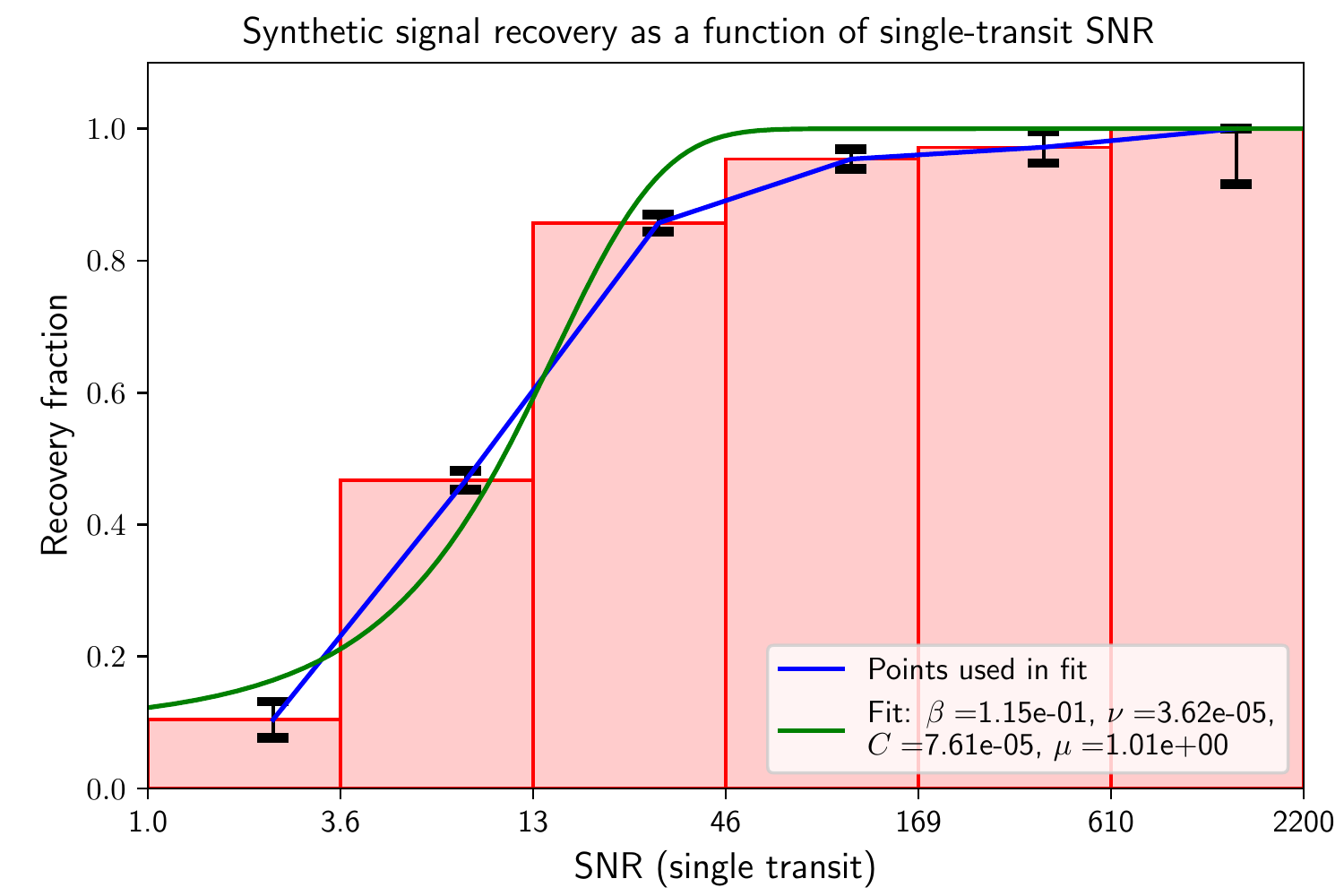}
\caption{Simulation recovery interpolated from $1 < R/R_\oplus \leq 15$ and $0.7 < P/\text{days} \leq 1000$ as a function of SNR, marginalized over orbital period.  The sigmoid (in green) was fit to the recovery fractions in each bin (connected in blue).  Fit values for coefficients are shown (see text).  Binning in SNR is logarithmically even.\label{fig.sigfit}}
\end{figure}

For any statistical use of the SATCHEL pipeline, a correction must be applied to account for the tendency of some types of signals to be missed more frequently than others.  In the case of the \kepler data, as we have seen in Sections~\ref{sec.simreco} and~\ref{sec.KOIreco}, the signals most frequently missed are shallow signals with low SNR, and the increase in recovery performance of the pipeline appears to be a smooth, monotonic trend as SNR increases.  There may also be one or more complex trends in pipeline performance with orbital period, but the trends implied for simulation and KOI recovery seem to contradict each other, as seen in Figs.~\ref{fig.recov-period-simsignals} and~\ref{fig.recov-period-KOIsignals}.  Due to the small number of KOIs in the longer period regime, we are currently unable to make strong statements about this contradiction or about pipeline completeness as a function of orbital period.  Constructing a completeness function that fully captures these intricacies is beyond the scope of this paper, as our primary goal herein is to offer SATCHEL as a generalized tool.  We thus refer the reader to Paper~II for a more detailed and specific discussion.

Instead, we include here an empirical treatment of completeness only for SNR, as an illustrative example.  Pipeline performance trends with SNR for simulations and KOIs, within the range of parameter space $1 \geq R/R_\oplus < 15$, are in robust agreement, as seen in Figs.~\ref{fig.recov-simsignals-radproxy} and~\ref{fig.recov-period-KOIsignals}.  With the same binning as shown in Fig.~\ref{fig.recov-period-simsignals}, recovery fractions for each bin were used as data points to fit a generalized logistic function of the form
\begin{equation}\label{eq.sigmoid}
    Q(\text{SNR}) = \frac{1}{\big[ 1 + C e^{-\beta(\text{SNR} - \mu)} \big]^{1/\nu}},
\end{equation}
with lower and upper asymptotes of 0 and 1, respectively.  A non-linear least squares fit (using the \texttt{curve\_fit} function of Python's \texttt{scipy.optimize} package) yields constants
\begin{alignat}{2}\label{eq.constants}
    C &= 7.61 \times 10^{-5}, &\qquad \beta &= 0.115, \\
    \mu &= 1.01, & \nu &= 3.62 \times 10^{-5}
\end{alignat}
for the fit, which is shown overlaid with recovery fraction for SNR bins in Figure~\ref{fig.sigfit}.  Marginalization over orbital period in Fig.~\ref{fig.sigfit} was done over the mean, rather than the sum; however, choice of marginalization does not noticeably alter either the recovery trend or the fit.

Taken as the pipeline completeness function, Eq.~\ref{eq.sigmoid} could be used to interpolate the pipeline completeness corrections for arbitrary SNR bins, or used in conjunction with other correctional factors (such as those addressing orbital geometry constraints and the \kepler observation window) \citep[see, e.g.,][]{bib.fore2016, bib.burk2015} to compute global occurrence rates for single exoplanet signals, given the number found in each bin after fully vetting the pipeline results, rejecting false positives, and correcting for reliability \citep{bib.brys2020a, bib.brys2020b}.  These tasks comprise the majority of Paper II.

\section{Summary and discussion}\label{sec.discussion}

We have presented SATCHEL, a pipeline capable of analyzing the 1,253,050 M-dwarf light curve classifications done by users of the Planet Hunters citizen science project.  PH users classified both real \kepler data and simulations constructed by injecting real \kepler light curves with synthetic signals corresponding to exoplanets occupying a representative range of parameter space in exoplanet radius and orbital period.  The classifications were comprised of either bounds of a marked region in the time-series data within which a user believed they saw a significant feature, or a non-marking, indicating that a user saw the time-series data but did not believe it contained any significant features.  The SATCHEL extracts the features marked by users, consolidates them into metafeatures where possible, and calculates, on a signal-by-signal basis, scores for each that represent agreement among users of that feature's significance, and by proxy the likelihood of that signal to be a legitimate exoplanet transit signal.  The score calculation is a form of user-weighted average, done first using weights assigned to users based on their performance at identifying synthetic exoplanet signals, with dummy weights for all users who did not see simulations, and done later using a coupled, iterative scheme which adjusts user weights for all users based on majority agreement with high-scoring signals and subsequent score recalculation.  The coupled adjustments continue until the weight and score distributions are both stable with a mean user weight of unity.  This process is computationally intensive, but necessary due to the significant fraction of our data that has been classified by users who saw no simulations.

We performed a recovery study, defining a known transit signal as recovered if it met matching criteria with a metafeature assigned a final score of 0.5 or higher by SATCHEL.  For signals corresponding to KOIs and synthetic exoplanets over 2~$R_\oplus$, about 72\% of KOI signals and about 86\% of synthetic signals were recovered.  At least one transit signal was recovered from every KOI, even those with the smallest SNR, corresponding to exoplanets down to 0.22~$R_\oplus$ in radius.  At this stage, the list of metafeatures that survive the cutoff is still highly contaminated by spurious features.  While choosing a more strict score cutoff decreases the raw number of these ``false positives'' (FPs) in the list, a preliminary study (see Appendix~\ref{sec.scorecutoff}) suggests that the {\it rate} of FPs is roughly uniform for any cutoff.  Filtering of FPs through multiple rounds of vetting specific to the study of long-period exoplanets and a thorough statistical treatment based on \citet{bib.fres2013} are done in Paper II.

The pipeline itself is best suited for offline analysis of a complete dataset.  Due to the computationally intensive coupling of the weighting and scoring, typical runtimes make real-time analysis unfeasible.  However, the coupled iteration is parallelized in several places, using Python's \texttt{multiprocessing} package.  Thus, an increase in core count would decrease runtime dramatically.  Further optimization of the pipeline itself may certainly be possible; both future and past versions will be archived on the SATCHEL GitHub repository.

The recovery studies for this pipeline, in particular, suggest that future citizen science projects with goals of a statistical nature would benefit from a larger number of simulations.  The needs will certainly vary with the project and dataset, but it seems reasonable to assert that some standard minimum number of injections should be constructed for each bin of physical parameter space, and that these bins should be chosen based on the binning most likely to be used for later analysis, led by contemporary understanding of the real distribution of signals across physical parameter space in nature.  In general, the complexity of this pipeline has been motivated by the large fraction of users who never interacted with simulations at all (see Figs.~\ref{fig.classsynthvsreal} and~\ref{fig.usersbyclass}), which could be avoided in future citizen science projects by either requiring users to undergo a short training period before exposure to real data or simply making the first subject displayed in the interface a simulation by default.  A bias may be introduced if the user knows to expect a signal to be present in the subject, but including simulations containing no injected signals, ``true negatives,'' could potentially mitigate such a bias.  With more simulations classified by a larger percentage of the user base, a more sophisticated weighting scheme may also be possible, in which reliability as a function of, e.g., varying signal depths could be calculated for individual users, rather than the global population \citep[e.g., the Space Warps Analysis Pipeline (SWAP)\footnote{On GitHub:  \href{https://github.com/zooniverse/swap}{https://github.com/zooniverse/swap}},][]{bib.mars2016}.

Maximizing user retention must be a high priority for future projects.  With the high degree of public attention that the PH project received came the side effect that many users visited the site only very briefly to classify one or two subjects (see Fig.~\ref{fig.usersbyclass}).  Studies of similar projects have shown that users who have classified more data generally perform better than brand new users \citep{bib.eisn2020b}.  Projects that are able to retain users for longer periods may have more options for methods of calculating and calibrating user weights \citep[S12,][]{bib.eisn2020b,bib.lint2008}.

It should be noted:  A citizen science project that advertises immediate contribution to real astronomy research should, ideally, be able to utilize participation from all volunteers.  This has been one of the cornerstone guiding principles in constructing the SATCHEL pipeline, though, at the inception of PH, it was impossible to foresee some of the challenges that would arise in the effort to remain faithful to the ideal.  We present this pipeline as a potential tool for analyzing other past and future PH classification data, or other crowd-sourced time-series data facing similarly complex needs.

\section*{Acknowledgements}

This work has been made possible by the participation of more than 30 000 volunteers in the Planet Hunters project, hosted on the Zooniverse platform.  The PH project utilizes data collected by the \kepler mission, which is funded by the NASA Science Mission directorate.  The authors would like to thank Geoff Clayton and Dan Foreman-Mackey for helpful discussions, as well as the reviewer for providing comments that have greatly improved the clarity and strength of the paper.  We would also like to acknowledge Satchel, the red lory, who provided the name of the pipeline and several very loud comments.

{\it Facilities:}  \kepler

{\it Software:}  matplotlib \citep{bib.hunt2007}, numpy \citep{bib.vand2011}, pandas \citep{reback2020pandas,mckinney-proc-scipy-2010}, multiprocess \citep{bib.mcke2012} \\

{\it Data availability:}  The data underlying this article were provided by the Zooniverse with permission.  Data will be shared on request to the corresponding author with permission of the Zooniverse.  The SATCHEL software is publicly available at \href{https://github.com/ejsafron/PH-pipeline}{https://github.com/ejsafron/PH-pipeline}.




\bibliographystyle{mnras}
\bibliography{refs} 




\appendix

\section{Optimization of pipeline design and hyperparameters}\label{sec.optimize}

Construction of the SATCHEL pipeline has involved some decisions that, while seemingly affecting only small detials, were non-trivial to make.  In an effort to make SATCHEL as useful as possible to other projects, the purpose of this Appendix is to clarify the tests and reasoning that led to those choices with a level of detail that would allow other projects to reproduce them and make the optimal decisions for their own purposes.

Section~\ref{sec.mfparam} discusses the optimization of hyperparameters used to construct the metafeature list and match user marks to synthetic signals in the calculation of user weight seeds.  Section~\ref{sec.decay} compares the decay function used in the user weight seed calculation to other possible methods.  And finally, Section~\ref{sec.scorecutoff} examines in detail how the post-pipeline choice of score cutoff affects recovery and false positive rates.

\subsection{Metafeature list hyperparameters}\label{sec.mfparam}

\begin{figure}
\centering
\includegraphics[width=1.0\columnwidth]{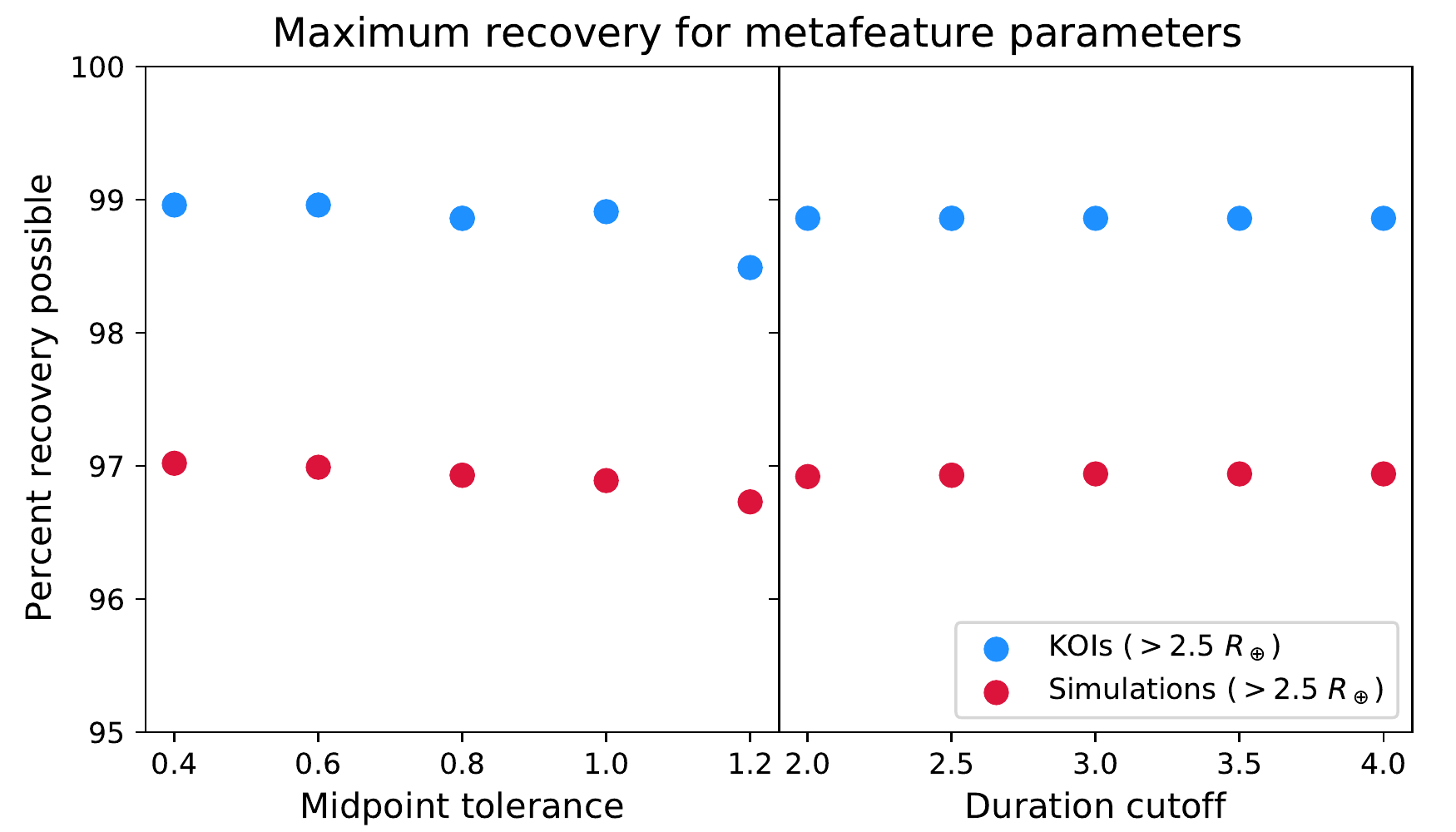}
\caption{Maximum possible signal recovery for KOIs and synthetic exoplanets based on metafeature lists constructed using a range of hyperparameters.  For those sets with varying midpoint tolerance (left panel), the duration cutoff was 2.5 days.  For sets with varying duration cutoff (right panel), the midpoint tolerance was 0.8 days.  Blue points correspond to KOIs, and red points correspond to simulations.\label{fig.max-recov}}
\end{figure}

\begin{figure}
\centering
\includegraphics[width=1.0\columnwidth]{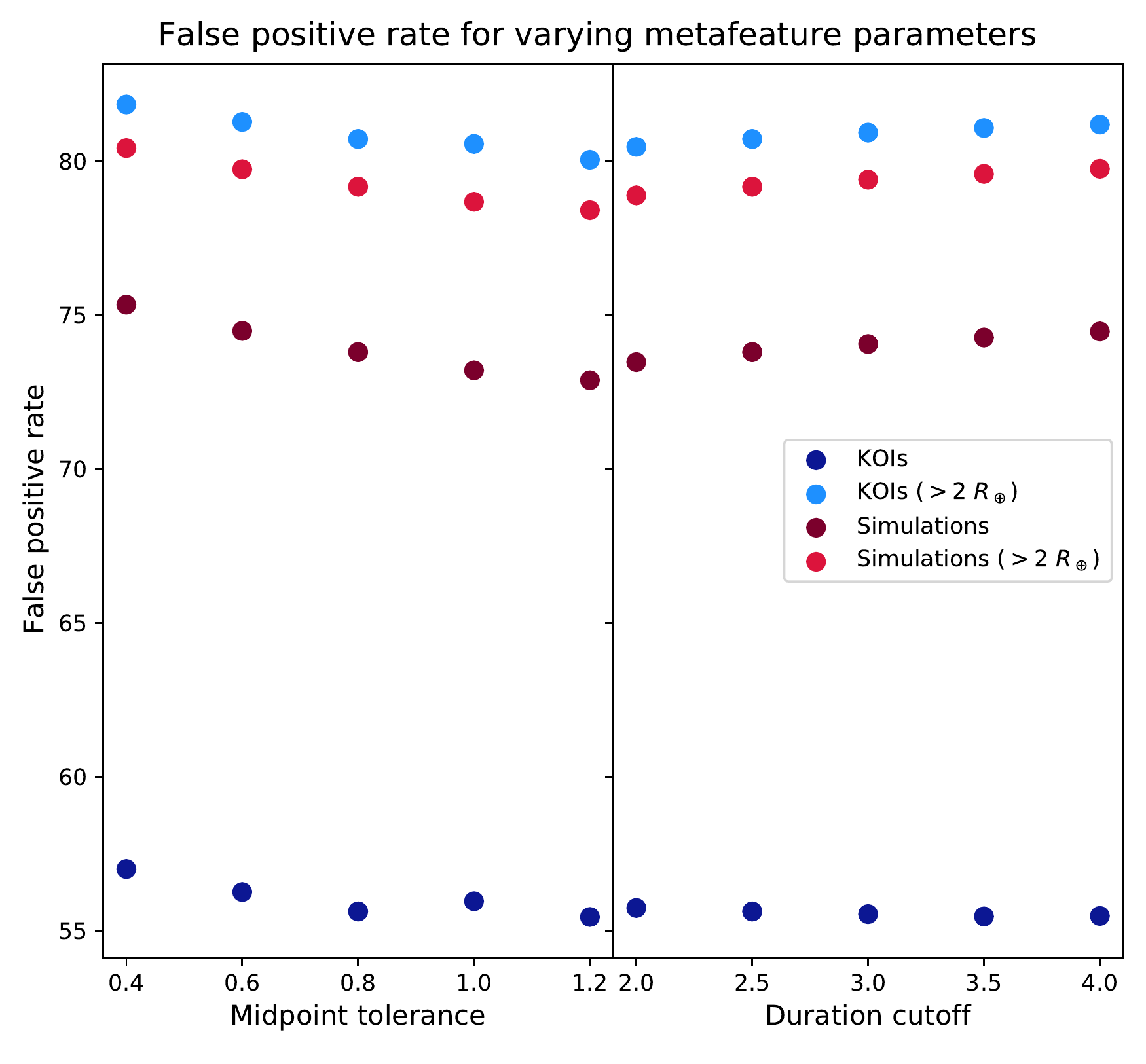}
\caption{False positive rates for unscored metafeatures in lists constructed using a range of hyperparameters.  For those sets with varying midpoint tolerance (left panel), the duration cutoff was 2.5 days.  For sets with varying duration cutoff (right panel), the midpoint tolerance was 0.8 days.\label{fig.fp-rates}}
\end{figure}

The metafeature list described in Sec.~\ref{sec.metafeatures} is constructed by a process that utilizes two hyperparameters:  a ``midpoint tolerance'' (\texttt{tolerance}), the largest acceptable difference in midpoints between two user marks indicating the same feature in a subject; and a ``duration cutoff'' (\texttt{cutoff}), the longest acceptable duration of a user mark that might indicate a significant feature.  Varying these two parameters does produce slightly different lists, and the metafeatures contained therein may be different in duration and/or midpoint, and more or fewer in number.

While the scoring process provides a useful way to filter out the bulk of metafeatures that likely do not correspond to real signals of interest, it remains true that some signals from both simulations and known KOIs will be lost in the low-scoring regime, simply because many people missed them.  In general, the maximum possible number of known signals recoverable from the PH data --- and the maximum number of spurious signals --- are obtained when no scoring whatsoever is done and the metafeature list is accepted in its entirety.  Since these recovery and ``false positive'' values depend only on the properties of the metafeatures, we use them to help determine the optimal set of hyperparameters with which to build the metafeature list.  That is, we choose the set of parameters that produces the set of metafeatures with the highest maximum possible recovery and the lowest corresponding false positive rate.  An equivalent way to describe this methodology is to say that we want the highest possible maximum recall and precision.

Thus far, we have implicitly defined a “positive” as any metafeature that passed a specific post-pipeline score cutoff threshold (recall, in Secs.~\ref{sec.KOIreco} and~\ref{sec.simreco}, we applied a score cutoff of 0.5).  For the purpose of this study, to examine the full metafeature list, we will allow the score floor to be zero, and we will explicitly define a ``true positive'' as any metafeature that corresponds to a known signal --- the transit of either a KOI or a synthetic exoplanet.  A ``false positive,'' then, is a metafeature that does not correspond to a KOI or synthetic exoplanet signal.  We limit this discussion to, for KOIs, subjects that correspond to only \kepler targets associated with KOIs, and for synthetic exoplanets, only simulation subjects.  That is, {\it we do not use the full set of metafeatures to calculate the false positive rate in either case.}  We consider this the most correct choice for several reasons:  %
\begin{itemize}
    \item No true negatives were included in construction of the simulations; every simulation did contain at least one synthetic transit.  To explore the truest possible comparison between the KOI and simulation false positive and recovery rates, we included in the KOI study only subjects displaying data from KOI hosts (though, for some of the longer-period KOIs, this does include subjects that contain no KOI transits).
    \item To label high-scoring, significant features in targets that are not KOI hosts as false positives, while technically correct as they are not KOI transits, is to imply that those features should be ignored.  This practice is incompatible with the spirit of the false positive study.
    \item KOI hosts (in particular, hosts with systems of multiple KOIs) have been more thoroughly explored for extra long-period planets \citep{bib.ueha2016} than targets hosting no known exoplanet candidates, and thus there is a smaller chance that a given false positive feature in a KOI host light curve is actually a signal from a real, yet undiscovered exoplanet.
    \item Users searching for exoplanet transits in the light curve subjects of KOI hosts are effectively, though unknowingly, {\it specifically searching for KOI transits}.  The same cannot be said of user classifications of data not corresponding to KOI hosts, therefore the definition of a false positive as ``a user mark or set of user marks indicating the transit of a KOI where no such transit truly occurred'' is not applicable.
\end{itemize}

\begin{figure}
\centering
\includegraphics[width=1.0\columnwidth]{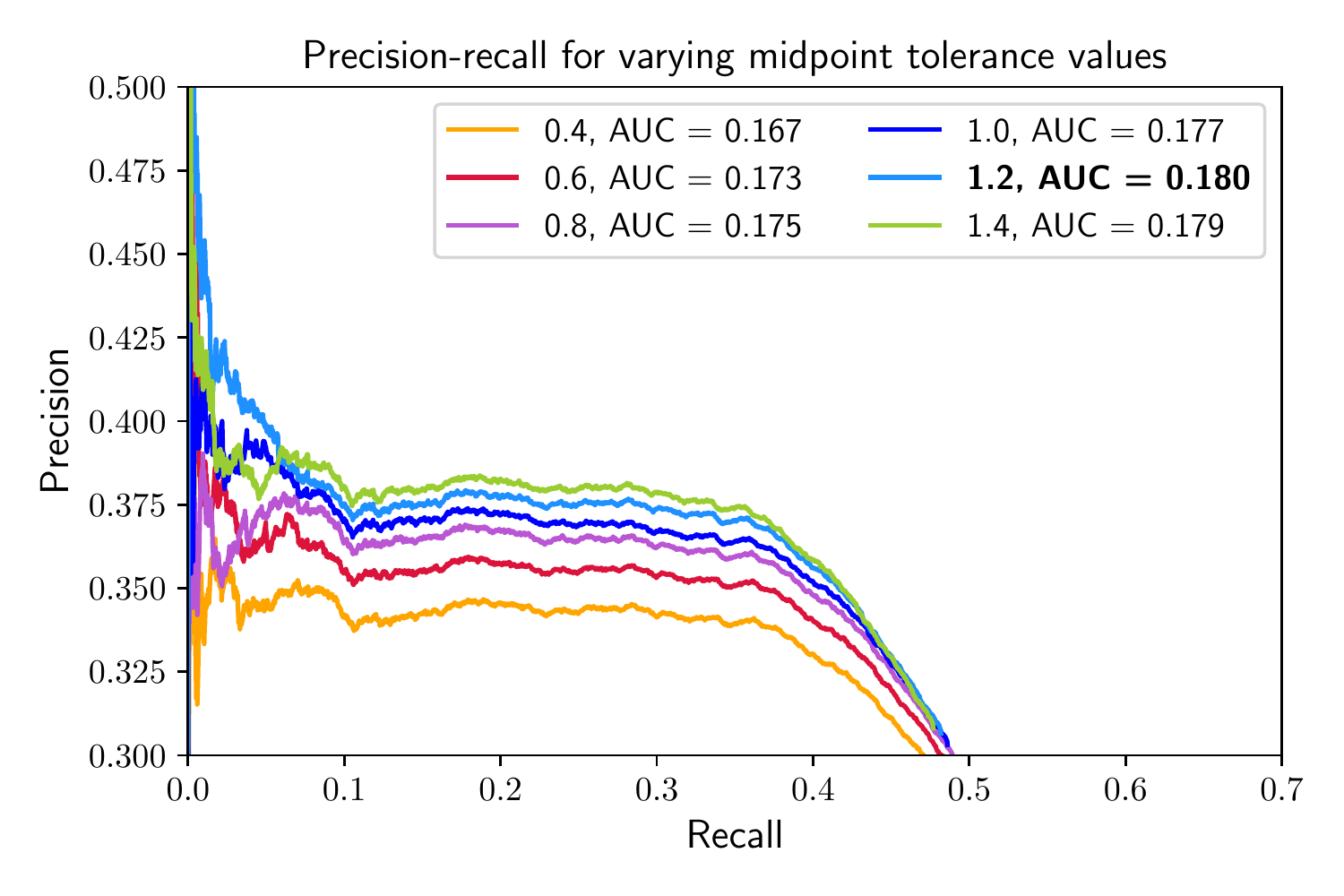}
\caption{Precision versus recall for SATCHEL runs with different midpoint tolerance values (duration cutoff set to 2.5 days for all).  Area under the curve (AUC) is given for each run, calculated with the trapezoidal rule via the \texttt{sklearn} Python package.  The maximum area under the curve (AUC) corresponds to a midpoint tolerance of 1.2 days.\label{fig.PRmidtol}}
\end{figure}

\begin{figure}
\centering
\includegraphics[width=1.0\columnwidth]{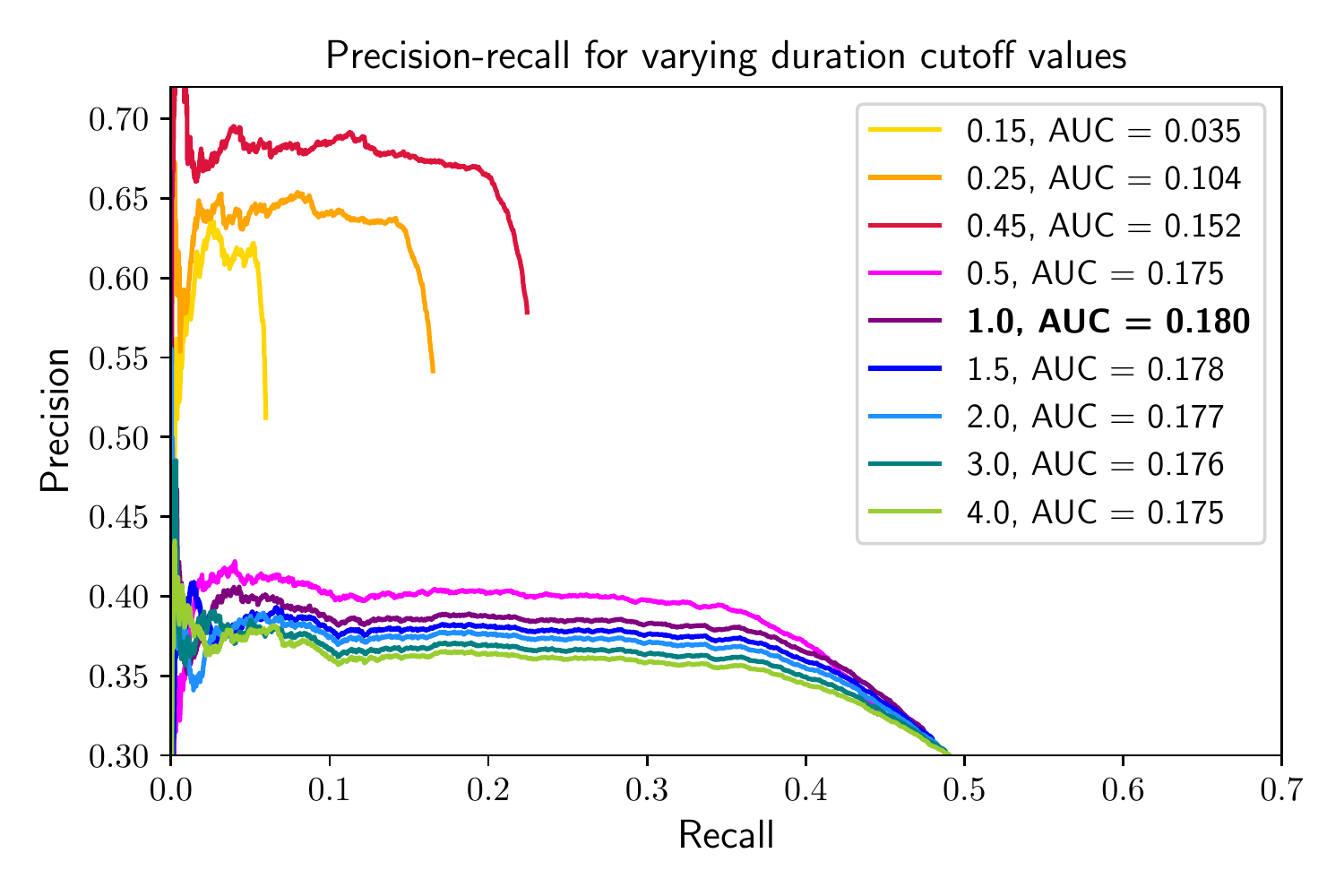}
\caption{Precision versus recall for SATCHEL runs with different duration cutoff values (midpoint tolerance set to 1.0 day for all).  AUC is given for each run, calculated with the trapezoidal rule via the \texttt{sklearn} Python package.  The maximum AUC corresponds to a duration cutoff value of 1.0 days.\label{fig.PRdurcut}}
\end{figure}

Metafeature lists were constructed and run through the rest of the SATCHEL pipeline using \texttt{tolerance} values from 0.4 to 1.4 days, in increments of 0.2 days (with \texttt{cutoff} $=2.5$ days), and \texttt{cutoff} values from 0.15 to 4.0 days (with \texttt{tolerance} $= 0.8$ days).  Figure~\ref{fig.max-recov} shows the percentage of signals corresponding to KOIs and synthetic exoplanets with radii$~>2~R_\oplus$ recoverable in metafeature lists with hyperparameters in these ranges.  Figure~\ref{fig.fp-rates} shows the false positive rates for the same metafeature sets.  Both the maximum possible recoveries and the false positive rates, for signals corresponding to objects$~>2~R_\oplus$ in radius, are roughly the same for KOIs and simulations, separated by only about two per cent.  This is a reassuring result, indicating that users were classifying the simulations the same way, qualitatively, that they were classifying the real data.  The significant difference in false positive rate for the whole set of KOIs and simulations may be attributable to the much larger population of KOIs with radii$~<2~R_\oplus$, and even occasionally$~<1~R_\oplus$, which was the cutoff for radii of the synthetic exoplanets.

The simulations and real data also display the same trends across varying hyperparameters:  A slight decrease in possible recovery and decrease in false positive rate with increasing midpoint tolerance, and stable recovery with increasing false positive rate for increasing duration cutoff.  A trend of higher possible recovery with increasing duration cutoff does appear for signals corresponding to objects with radii$~<2~R_\oplus$, but, as we do not expect the pipeline to be even approximately complete in this regime, these objects are not the focus of our study and we instead prioritize a lower false positive rate in our choice of cutoff.  

Figures~\ref{fig.PRmidtol} and \ref{fig.PRdurcut} show the precision as a function of recall for varying midpoint tolerance and duration cutoff values, respectively.  For varying midpoint tolerance, maximum recall and average precision both increase with midpoint tolerance up to about 1.0 day, after which average precision continues to rise but maximum recall falls.  Beyond a midpoint tolerance of 1.4 days, average precision also falls.  The maximum AUC corresponds to a midpoint tolerance of 1.2 days.  For varying duration cutoff, we find the highest maximum recall for the largest, most lenient values.  As the duration cutoff decreases, the maximum recall decreases and average precision increases.  Average precision rises steeply with duration cutoffs less than 0.5 days, but maximum recall decreases sharply as well.  There is a significant change in the behavior of the PR curve between duration cutoffs of 0.45 and 0.5 days.  While we are uncertain what the reason for this is, we offer that it may be due to a drastic change in behavior of PH volunteers at marking widths of 0.5 days; for instance, the default marking width in the PH interface may have been 0.5 days.  The maximum AUC corresponds to a duration cutoff value of 1.0 days.

Thus, we choose a midpoint tolerance of 1.2~days and a duration cutoff of 1.0~day.

\subsection{User weight decay function}\label{sec.decay}

\begin{figure}
\centering
\includegraphics[width=1.0\columnwidth]{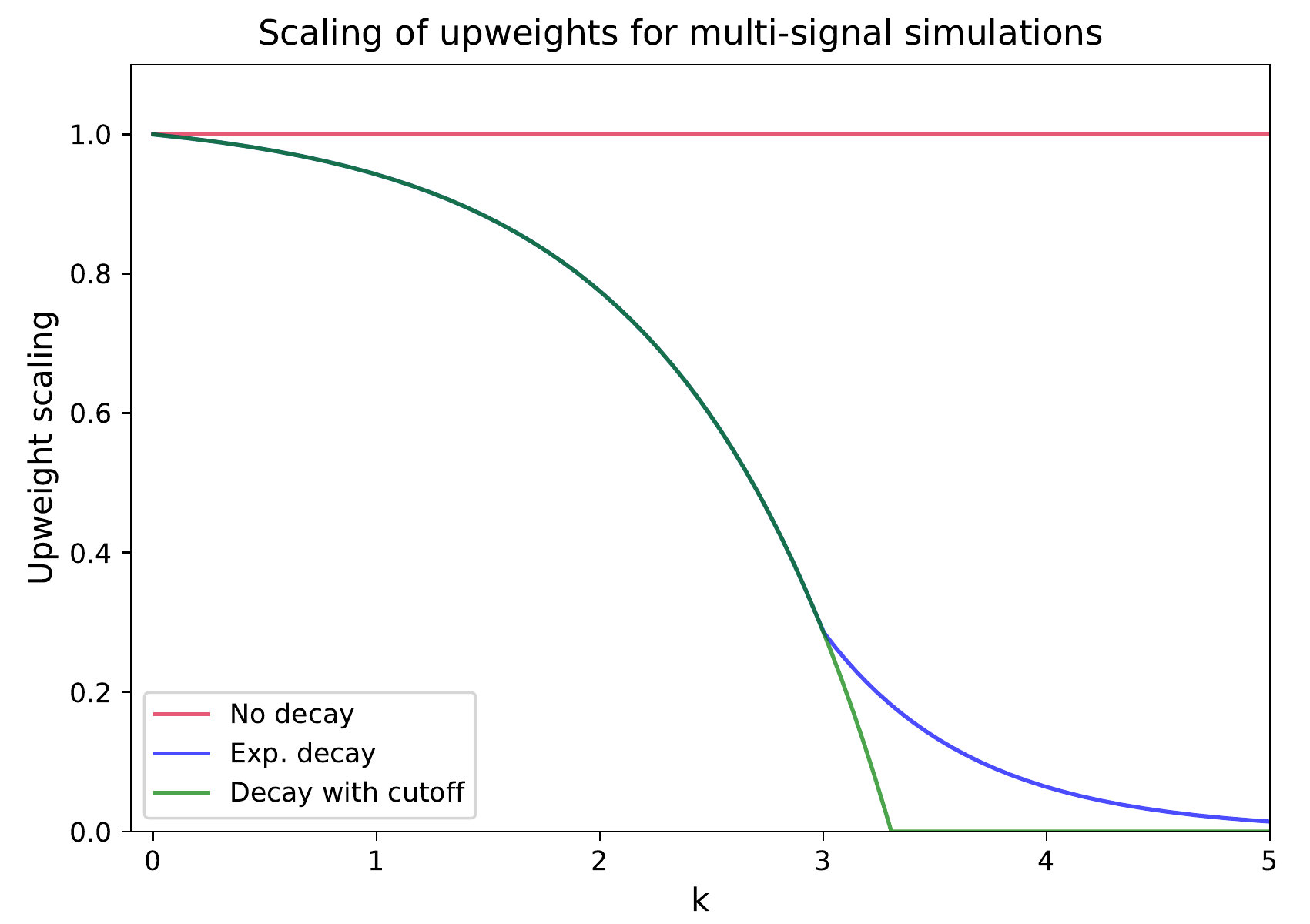}
\caption{Decay functions tested for upweighting scaling, where $[k]$ is the index of the user mark (if a single user has made $k+1$ correct marks on a single lightcurve), sorted in order of highest overlap fraction.  Colors are consistent with the decay functions used to produce the user weight distributions shown in Fig.~\ref{fig.decay-hist}.  The function with an exponential tail (blue) is the decay function used throughout the main body of this work.  See Sec.~\ref{sec.upweighting}.\label{fig.decay-fns}}
\end{figure}

\begin{figure}
\centering
\includegraphics[width=1.0\columnwidth]{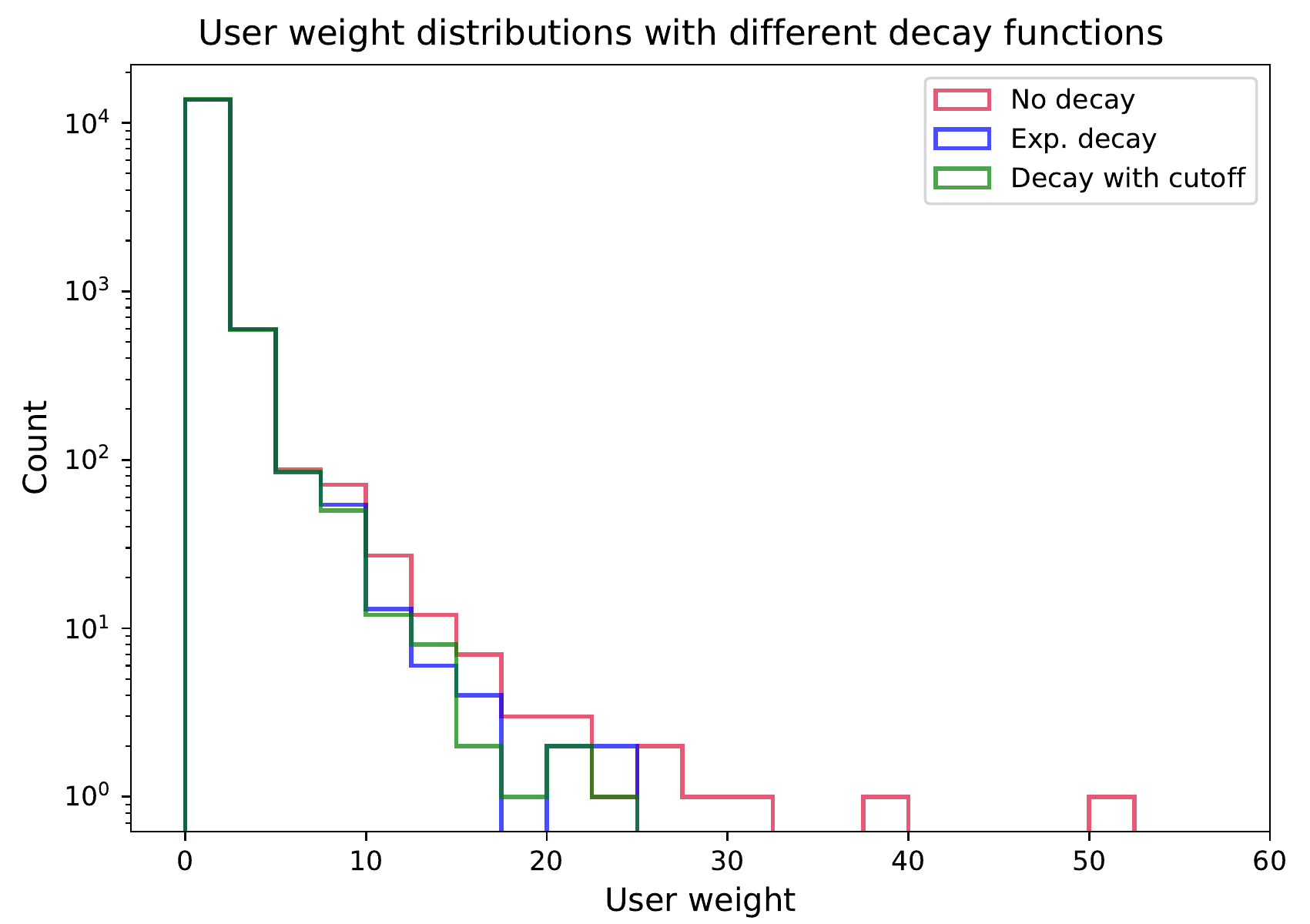}
\caption{User weight distributions calculated from the decay functions given in Sec.~\ref{sec.decay} and shown in Fig.~\ref{fig.decay-fns}.  The method for calculating user weights in all cases is that described in Sec.~\ref{sec.weighting}.\label{fig.decay-hist}}
\end{figure}

The need for a decay function in calculation of the user weight seeds was not immediately apparent in the initial design of the PH pipeline.  No decay function was implemented in previous PH studies, including S12, nor is any decay function used in the single-calculation user weighting process of the PH {\it TESS} pipeline.

In the case of the \kepler PH data, the decay function was conceptually motivated by examination of the distribution of user weight seeds.  Prior to inclusion of the decay function, the weights of some users were inflated to many times the mean weight, as shown by the red distribution in Figure~\ref{fig.decay-hist}.  With a mean weight of 1.0 after normalization, the highest weight was over 50 units in magnitude, and belonged to a user who had classified only three subjects, all of which were simulations and one of which contained seven shallow transits of a $1.5~R_\oplus$ synthetic exoplanet.  While, individually, these seven transits would all be equally difficult to find, the strictly periodic placement of them in a light curve subject decreases that difficulty after the first two or three signals found.  That is, after finding two transits, a user knows exactly where to look to find the next.  Thus, it seemed unfair to award the full raw upweight for each of a large number of signals identified in a single subject.

Rather, it was determined that a more reasonable approach would be to award full or very high upweights only to the few most accurately identified synthetic signals in a subject, with any others identified being given decreased amounts of upweight, until eventually the user is being awarded very little or no upweight for further signals identified.  A function consistent with this behavior was constructed, and a few variants devised.  Particularly, we tested a function with an inverse exponential tail as given in Equation~\ref{eq.decay}, which awards a low but non-zero fraction of the raw upweight for any synthetic signal identified past the fourth, and a similar function with a strict cutoff,
\begin{equation}\label{eq.decay-cutoff}
    f_\text{decay}(k) = \begin{cases}
        g(k) &\quad\text{if~}k \le 3, \\
        0 &\quad\text{if~}k>3,\\
        \end{cases}
\end{equation}
(where $g(k)$ is again identical to Eq.~\ref{eq.decay-inner}), which awards no upweight at all past the fourth synthetic signal identified.  These two functions are shown in blue and green in Figure~\ref{fig.decay-fns}, and their effect on the user weight seed distribution is shown in the same respective colors in Fig.~\ref{fig.decay-hist}.  Both functions significantly narrow the spread of the distribution toward large user weights, and both completely remedy the issue of the inflated outliers.  We chose to implement Eq.~\ref{eq.decay}, with the inverse exponential tail, because we deemed it more correct in principle to reward users for marking {\it all} significant features in a subject, even if that reward must be small.

The inner function $g(k)$, of the general form $a + b~e^{c k}$, contains three coefficients $a$, $b$, and $c$, which can be adjusted or optimized.  For the PH data, the user weight distributions resulting from implementation of both Eqs.~\ref{eq.decay} and~\ref{eq.decay-cutoff} were robust under slight variations of all coefficients, and, further, the final score distribution was robust under slight changes in the user weight seed distribution, so we did not feel that optimizing the coefficients was a high priority.  It is possible that data from other projects could be more sensitive to the choice of coefficients, in which case optimization may be necessary.  On the other hand, however, for projects in which the signals of interest are not periodic or patterned in nature, the decay function may be done away with entirely, or set simply to unity.

\subsection{Score cutoff}\label{sec.scorecutoff}

\begin{figure}
\centering
\includegraphics[width=1.0\columnwidth]{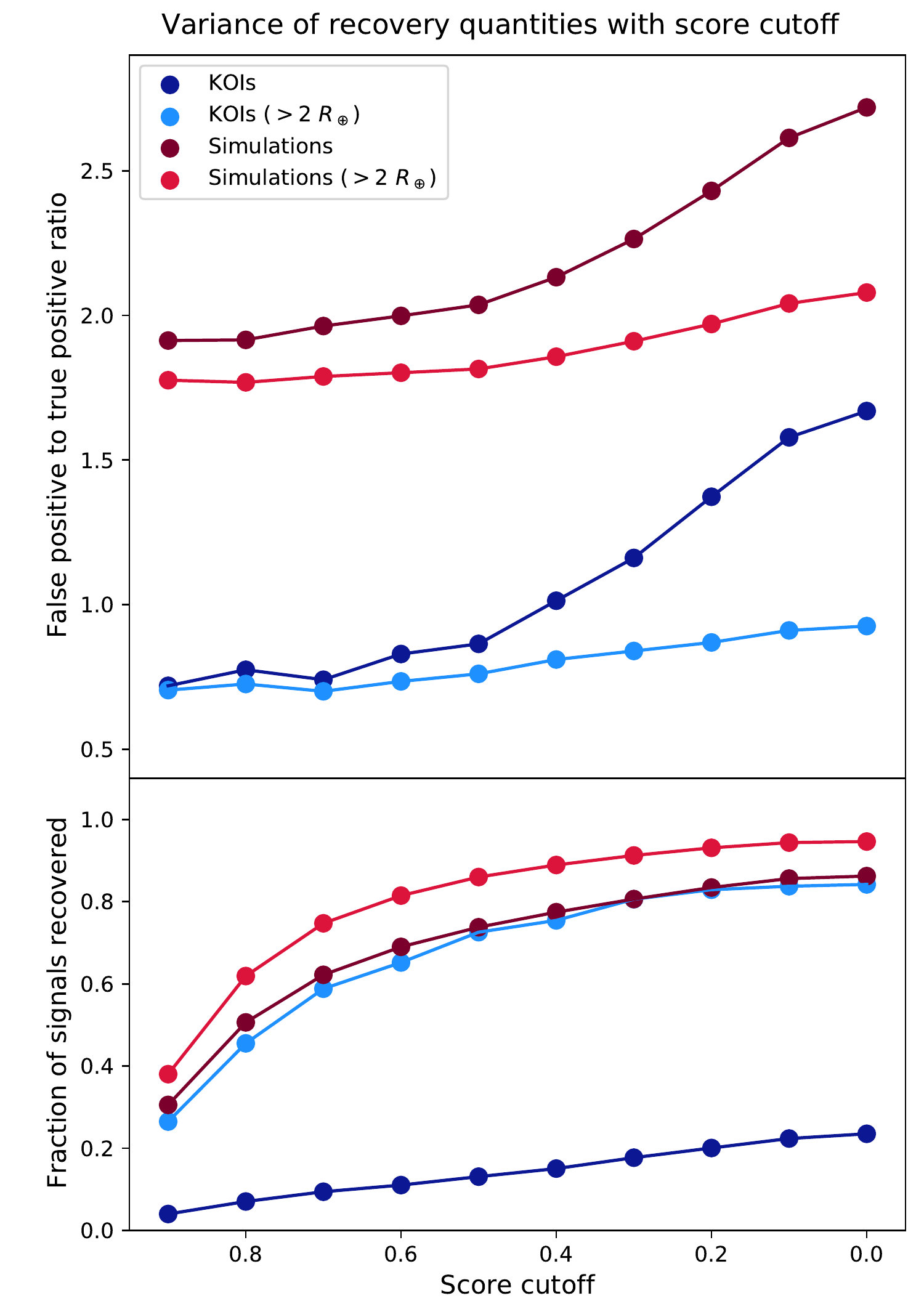}
\caption{Bottom:  Fraction of signals recovered with score cutoffs ranging from extremely strict (left end of the horizontal axis) to none (right end), for simulations containing synthetic exoplanets with radii $>2~R_\oplus$ (red), all simulations (dark red), KOIs with radii $>2~R_\oplus$ (blue), and all KOIs (dark blue).  Top:  Ratio of false positives to true positives.  Please see text for clarification on the definition of a false positive in this context.  No artificial offset has been added between the simulations and the KOIs.  Signal matches were generated by a pipeline run with global hyperparameters of midpoint tolerance $= 1.0$ days and duration cutoff $= 2.5$ days.\label{fig.score-cutoff}}
\end{figure}

The purpose of the score cutoff, in practice, is to limit any further vetting of metafeatures to only the most promising subset.  Much like the hyperparameters discussed in Section~\ref{sec.mfparam}, choosing the optimal score cutoff with which to filter results from the pipeline is a balancing act that may vary from project to project.  Generally, we seek to maximize the recovery of truly interesting signals while minimizing contamination by spurious features.

For this study, our definitions of true and false positives will be fundamentally identical to those of Appendix~\ref{sec.mfparam}, in that a ``positive'' shall be any metafeature that passes a specific score cutoff threshold.  We investigated the effect of applying different score thresholds on the sample, by recalculating recovery and false positive rates with score floors varying from 0.0 (no floor) to 0.9 (the most restrictive), the results of which are summarized in Figure~\ref{fig.score-cutoff}.  The bottom panel, displaying recovery fraction as a function of score cutoff for both KOIs and simulations (and subsets of both with radii $> 2~R_\oplus$) shows a steep rise as the score cutoff is made less strict, dropping from values of 0.9 to 0.8 and 0.7.  At the point where all metafeatures with scores $> 0.6$ are included, over 65\% of signals produced by KOIs larger than $2~R_\oplus$ are recovered and over 80\% of signals produced by the same size subset of synthetic exoplanets.  Dropping the score cutoff to 0.5 increases signal recovery for these objects to over 70\% for KOIs, and over 85\% for simulations.  Below about 0.5, however, the increase in recovery rate is slower, leveling out significantly toward the lowest score values.  The outlier case is that of the full set of KOIs.  The presence there of very small exoplanets, much smaller than the smallest of the synthetic exoplanets, keeps the signal recovery rate extremely low even at the most lenient of score cutoffs.  While recovery does increase as the score cutoff decreases, the increase is almost perfectly linear, unlike that of the other subsets examined.

Of course, as the score cutoff decreases, the false positive rate also increases.  As shown in the top panel of Fig.~\ref{fig.score-cutoff}, around the same score cutoff value of 0.5, the ratio of false positives to true positives begins to rise at a higher rate, particularly for the full sets of KOIs and simulations.  This is interesting, in itself --- it may indicated that, in the absence of obvious signals, users are more likely to mark ambiguous features.  Or, conversely, that in the presence of obvious signals, users may ignore ambiguous features, potentially missing some transit signals that are real, but much shallower.  For the subsets of KOIs and synthetic exoplanets larger than $2~R_\oplus$, the false positive to true positive ratio does increase more steeply below a score cutoff of 0.5, but the difference is more subtle.

The combination of the upper and lower panels of Fig.~\ref{fig.score-cutoff} suggests that a score cutoff near the middle is most practical.  In the more linear low-scoring regime of the bottom panel, where the signal recovery fraction is increasing shallowly but steadily, the false positive to true positive ratios are also increasing in a fairly linear fashion.  The proportionality between these increases begins to deviate in a significant way at a score cutoff of 0.5.  Thus, we choose 0.5 as the working score cutoff for our analyses in both Section~\ref{sec.recovery} and Paper II.

A last oddity to note is that the false positive to true positive ratios for simulations appear to be twice as high as their KOI counterparts, in Fig.~\ref{fig.score-cutoff}.  As far as the authors are aware, there was no difference in user methodology when classifying simulations or KOIs, and no significant qualitative differences in either the transit signals or the background data.  The source of this offset may be worth exploring in the future.


\bsp	
\label{lastpage}
\end{document}